\documentclass[10pt,aps,prb,twocolumn,floatfix,floats,showpacs,superscriptaddress,raggedbottom]{revtex4-1}
\usepackage{graphicx,latexsym}
\usepackage{dcolumn}
\usepackage{amssymb,amsmath,bm}

\pdfoutput=1
\begin{document}

\title{Time-dependent transport of electrons through a photon cavity}

\author{Vidar Gudmundsson}
\email{vidar@hi.is}
\author{Olafur Jonasson}
\affiliation{Science Institute, University of Iceland, Dunhaga 3,
        IS-107 Reykjavik, Iceland}
\author{Chi-Shung Tang}
\email{cstang@nuu.edu.tw}
\affiliation{Department of Mechanical Engineering,
        National United University,
        1, Lienda, Miaoli 36003, Taiwan}
\author{Hsi-Sheng Goan}
\email{goan@phys.ntu.edu.tw}
\affiliation{Department of Physics and Center for Theoretical Sciences, 
             National Taiwan University, Taipei 10617, Taiwan}
\affiliation{Center for Quantum Science and Engineering, 
             National Taiwan University, Taipei 10617, Taiwan}
\author{Andrei Manolescu}
\affiliation{Reykjavik University, School of Science and Engineering,
        Menntavegur 1, IS-101 Reykjavik, Iceland}

%

\begin{abstract}
We use a non-Markovian master equation to describe the transport of Coulomb
interacting electrons through an electromagnetic cavity with one quantized
photon mode. The central system is a finite parabolic quantum wire that is
coupled weakly to external parabolic quasi-one-dimensional leads at $t=0$.
With a stepwise introduction of complexity to the description of the system
and a corresponding stepwise truncation of the ensuing many-body spaces we
are able to describe the time-dependent transport of Coulomb-interacting
electrons through a geometrically complex central system. 
We take into account the full electromagnetic interaction of electrons and 
cavity photons without resorting to the rotating wave approximation or 
reduction of the electron states to two levels. 
We observe that the number of initial cavity photons and their 
polarization can have important effects on the transport properties of the
system. The quasiparticles formed in the central system have a lifetime
limited by the coupling to the leads and radiation processes active on
a much longer timescale.   

\end{abstract}


\maketitle
\section{introduction}
During the past decade there has been increasing interest in
exploring time-dependent quantum transport through open mesoscopic
systems in a strong system-lead coupling
regime.\cite{Tang99:1830,Tang03:205324,Zhou05:6663,Zeb08:165420,Lin08:245312,Torfason09:195322,Kienle09:026601}
Utilizing the tunable dynamic response of transient
time-dependent transport enables development of switchable
mesoscale electronic devices, in which the interplay of the
mesoscopic system with external perturbations plays an important
role.\cite{Myohanen09:115107,Stefanucci10:115446,Abdullah10:195325,Tahir10:195444,Chen11:115439}

In the weak system-lead coupling regime, the wide-band and the
Markovian approximation are usually employed, by neglecting the energy
dependence of the electron tunneling rate, as well as memory
effects in the system, respectively.\cite{Gurvitz96:15932,Kampen01:00,Harbola06:235309}  
It is assumed that the correlation time of the electrons in the leads
is much shorter than the typical response time of the central
system.  However, transient transport properties,
which are intrinsically linked to coherence and relaxation dynamics, cannot 
generally be described in the Markovian limit.  One has to take into
account the energy-dependent spectral density in the leads, and an
accurate numerical method for such a nonequilibrium transient
time-dependent transport is desirable.  A non-Markovian
density-matrix formalism involving the energy dependent coupling
elements should be considered based on the generalized master
equation
(GME).\cite{Braggio06:026805,Emary07:161404,Bednorz08:206803,Moldoveanu09:073019,Gudmundsson09:113007,Vaz08:012012}
How to appropriately describe the carrier quantum dynamics under
nonequilibrium conditions in realistic device geometries is a
challenging problem.\cite{Abdullah10:195325,Gudmundsson09:113007,Gudmundsson10:205319}

More recently, manipulation of the electron-photon coupled quantum
systems in an electromagnetic cavity has become one of the key
issues to be implemented in quantum information processing
devices.  Utilizing the giant dipole moments of intersubband
transitions in quantum wells\cite{Helm00:01,Gabbay11:203103} enables researchers to
reach the ultrastrong electron-photon coupling
regime.\cite{Ciuti05:115303,Devoret07:767,Abdumalikov08:180502}  In this
regime, the dynamical electron-photon coupling mechanism has to be
explored beyond the wide-band and rotating-wave
approximations.\cite{Zela97:106,Sornborger04:052315,Irish07:173601}
Nevertheless, the dynamical time-dependent transport of Coulomb
interacting electrons in a specified geometry through an
electromagnetic cavity with quantized photon modes remains
unexplored.

In the present work, we explore the electronic transient transport
dynamics of an open quantum wire placed in a linearly polarized 
electromagntic field created in a cavity.\cite{Delbecq11:01,Frey11:01}
The wire is contacted to
two quasi-one-dimensional (Q1D) semi-infinite leads. 
A bias voltage is suddenly switched on between the leads, along the
wire. The whole structure is considered to be placed in a 
parpendicular homogeneos external magnetic field. 
We use the Nakajima-Zwanzig (N-Z) formalism to project the time
evolution of the system onto the Hilbert space of the 
central element (the short wire) by taking trace with 
respect to the operators in the leads.\cite{Nakajima58:948,Zwanzig60:1338} 
The transient transport properties will be explored by calculating the
time-dependent total mean number of electrons and photons, and the
time-dependent total charge current from the left (L) lead to the
right (R) lead.

The paper is organized as follows. In Sec.\ II, we shall describe
our theoretical model including the electron system in an
electromagnetic cavity by coupling a many-level electron system with
photons using a full photon energy spectrum of a single cavity mode
in the Fock space.  The N-Z framework is utilized to describe the system-lead
coupling.\cite{Nakajima58:948,Zwanzig60:1338} In section III we
investigate the dynamical transient transport properties.
Concluding remarks will be presented in Sec.\ IV.

\section{Model}
We start by describing a Coulomb interacting electron system in a 
closed finite system in an electromagnetic cavity with quantized photon modes
of two different polarizations. Later we shall open the system
by coupling it to external Q1D leads in order to allow for transport of electrons through it. 
In the closed system the number of electrons is constant, but in order to accomplish
the opening up of the system we need to calculate its equilibrium properties for a
varible number of electrons. In the equilibrium
and transport calculation to follow $N_\mathrm{SES}$ will represent the number of Single-Electron 
States (SESs) used to build the Many-Electron States (MESs) of the 
system.\cite{Moldoveanu09:073019,Gudmundsson09:113007}

\subsection{Electron system in an electromagnetic cavity}
The central element, a short wire placed  
in the $x$-$y$-plane is in an external homogeneous classical magnetic field constant in time,
$\mathbf{B}=B\hat{\mathbf{z}}$, introduced by the vector potential $\mathbf{A}_{\mathrm{ext}}$, 
and cavity fields represented by a quantized vector field $\mathbf{A}$.
The many-body Hamiltonian of this closed system is
\begin{align}
      H&_0=H_{\mathrm{Coul}}+H_{\mathrm{EM}}\nonumber\\
      &+\int d{\mathbf{r}}\; \psi^\dagger\left\{\frac{1}{2m^*}\left(
      \mathbf{p}+\frac{e}{c}\left[\mathbf{A}_{\mathrm{ext}}+\mathbf{A}\right] 
      \right)^2+\frac{1}{2}m^*\Omega_0^2y^2\right\}\psi \nonumber\\
      &=\int d{\mathbf{r}}\; \psi^\dagger\left\{\frac{1}{2m^*}\left(
      \mathbf{p}+\frac{e}{c}\mathbf{A}_{\mathrm{ext}}\right)^2
      +\frac{1}{2}m^*\Omega_0^2y^2 \right\}\psi \nonumber\\
      &+H_{\mathrm{Coul}}+H_{\mathrm{EM}}
      -\frac{1}{c}\int d{\mathbf{r}}\;\mathbf{j}\cdot\mathbf{A}
      -\frac{e}{2m^*c^2}\int d{\mathbf{r}}\;\rho A^2\nonumber\\
      &=\int d{\mathbf{r}}\;\psi^\dagger\left\{\frac{{\pi}^2}{2m^*}+\frac{1}{2}m^*\Omega_0^2y^2\right\}\psi
      +H_{\mathrm{Coul}}+H_{\mathrm{EM}}\nonumber\\
      &-\frac{1}{c}\int d{\mathbf{r}}\;\mathbf{j}\cdot\mathbf{A}
      -\frac{e}{2m^*c^2}\int d{\mathbf{r}}\;\rho A^2,
\label{Full-e-EM-I}
\end{align}
with $H_{\mathrm{Coul}}$ the Hamiltonian for the Coulomb interaction of the electrons 
and $H_{\mathrm{EM}}$ the Hamiltonian of the cavity photons to be introduced later and 
the charge current density and charge density
\begin{equation}
      \mathbf{j} = -\frac{e}{2m^*}\left\{\psi^\dagger\left({\bm{\pi}}\psi\right)
                                +\left({\bm{\pi}}^*\psi^\dagger\right)\psi\right\},\quad
      \rho = -e\psi^\dagger\psi, 
\end{equation}
where
\begin{equation}
      {\bm{\pi}}=\left(\mathbf{p}+\frac{e}{c}\mathbf{A}_{\mathrm{ext}}\right).
\end{equation}
The static magnetic field $\mathbf{B}=\bm{\nabla}\times\mathbf{A}_{\mathrm{ext}}$ together with the 
parabolic confinement of the q-1DES introduce a characteristic length 
$a_w=\sqrt{\hbar /(m^*\Omega_w)}$, with $\Omega_w^2=\omega_c^2+\Omega_0^2$ and
the cyclotron frequency $\omega_c=eB/(m^*c)$. The frequency $\Omega_0$ characterizes the
strength of the electron confinement in the $y$-direction. The finite parabolic quantum
wire has length $L_x$ and hard walls at $x=\pm L_x/2$. 

In terms of creation and annihilation operators the Hamiltonian of the
closed system takes the form
\begin{align}
        H&_0=H_\mathrm{e}+H_\mathrm{EM}+H_\mathrm{Coul}+H_\mathrm{e-EM}\nonumber\\
        &=\sum_i E_id_i^{\dagger}d_i + \hbar\omega a^{\dagger}a +
        \frac{1}{2}\sum_{ijrs}\langle ij|V_{\mathrm{Coul}}|rs\rangle 
        d_i^{\dagger}d_j^{\dagger}d_sd_r \nonumber\\ \label{H-e-EM}
        &+{\cal E}_\mathrm{c}\sum_{ij}d_i^{\dagger}d_j\; g_{ij}\left\{a + a^\dagger\right\}\\
        &+{\cal E}_\mathrm{c}\left(\frac{{\cal E}_\mathrm{c}}{\hbar\Omega_w}\right)
        \sum_{i}d_i^{\dagger}d_i\left\{\left( a^\dagger a+\frac{1}{2}\right) +
        \frac{1}{2}\left( aa+a^\dagger a^\dagger\right)\right\}{\nonumber},
\end{align}
where the single-electron states (SESs) of the closed system are labeled with Latin 
indices, $\{i,j,r,s\}$, $a$ is the destruction operator of one quantum of the single-mode
cavity field with frequency $\omega$, and $d_i$ is an annihilation operator of 
the non-interacting single-electron state $|i\rangle$ with energy $E_i$.
  
The electromagnetic cavity is a rectangular box $(x,y,z)\in\{[-a_\mathrm{c}/2,a_\mathrm{c}/2]
\times [-a_\mathrm{c}/2,a_\mathrm{c}/2]\times [-d_\mathrm{c}/2,d_\mathrm{c}/2]\}$ with the 
finite quantum wire centered in the $z=0$ plane. 
The polarization of the electric field can be chosen
along the transport direction, $x$, or perpendicular to it by selecting
the TE$_{011}$ or the TE$_{101}$ mode, respectively. In the Coulomb gauge   
the vector field is then
\begin{equation}
      \mathbf{A}(\mathbf{r})=\left({\hat{\mathbf{e}}_x\atop \hat{\mathbf{e}}_y}\right)
      {\cal A}\left\{a+a^{\dagger}\right\}
      \left({\cos{\left(\frac{\pi x}{a_\mathrm{c}}\right)}\atop\cos{\left(\frac{\pi y}{a_\mathrm{c}}\right)}} \right)
      \cos{\left(\frac{\pi z}{d_\mathrm{c}}\right)}, 
\label{Cav-A}
\end{equation}
with the upper equation representing the TE$_{011}$ mode and the lower one for TE$_{101}$.

The effective dimensionless coupling tensor of the electrons to the cavity mode 
due to the linear term in $\mathbf{A}$ in Eq.\ (\ref{Full-e-EM-I}) is
\begin{align}
\label{g_ab}
      g_{ij} = \frac{a_w}{2\hbar}\int d{\mathbf{r}}\; 
      [&\psi_i^*(\mathbf{r})\left\{\left(\hat{\mathbf{e}}\cdot\bm{\pi}\right)
      \psi_j(\mathbf{r})\right\}\\ \nonumber
      &+\left\{\left(\hat{\mathbf{e}}\cdot\bm{\pi}\right)
      \psi_i(\mathbf{r})\right\}^*\psi_j(\mathbf{r}) ],
\end{align}
since we have introduced the characteristic energy scale 
${\cal E}_\mathrm{c}=e{\cal A}\Omega_wa_w/c = g^\mathrm{EM}$ for the electron-cavity photon coupling.  
In the calculation of the energy spectrum of the electron-photon Hamiltonian
(\ref{H-e-EM}) we will retain all resonant and antiresonant terms 
in the photon creation and annihilation operators and not use
the rotating wave approximation, but in the calculations of the electron-photon
coupling tensor (\ref{g_ab}) we assume $a_w,L_x << a_\mathrm{c}$ and approximate
$\cos(\pi\{x,y\}/a_\mathrm{c})~\sim 1$ in Eq.\ (\ref{Cav-A}) for the cavity vector field $\mathbf{A}$. 

After the construction of the Fock space $\{|\alpha\rangle\}$ with the SESs 
that will later be deemed as relevant to the electron transport an exact numerical
diagonalization is used to obtain the Coulomb interacting MESs
$\{|\mu )\}$ with the energy spectrum $\tilde{E}_{\mu}$ and the unitary 
transformation\cite{Moldoveanu10:155442,Gudmundsson10:205319}
\begin{equation}
      |\mu ) = \sum_{\alpha}{\cal V}_{\mu\alpha}|\alpha\rangle .
\label{V}
\end{equation}
The total Hamiltonian of the electrons and the cavity fields can then be written
in terms of the interacting MESs
\begin{align}
      H_0&=\sum_\mu |\mu )\tilde{E}_\mu(\mu| + \hbar\omega a^{\dagger}a \nonumber\\ 
      &+ {\cal E}_\mathrm{c}\sum_{\mu\nu ij}|\mu )
      \langle\mu |{\cal V}^+ d_i^{\dagger}d_j{\cal V}|\nu\rangle (\nu |\;
      g_{ij}\left\{a+a^\dagger\right\} \nonumber\\
      &+{\cal E}_\mathrm{c}\left(\frac{{\cal E}_\mathrm{c}}{\hbar\Omega_w}\right)
      \sum_{\mu\nu i}
      |\mu )\langle\mu |{\cal V}^+ d_i^{\dagger}d_i{\cal V}|\nu\rangle (\nu | \nonumber\\   
      &\quad\quad\quad\quad\quad\quad\quad\left\{\left( a^\dagger a+\frac{1}{2}\right) +
      \frac{1}{2}\left( aa+a^\dagger a^\dagger\right)\right\}
\label{H-e-VEMV}
\end{align}
and its energy spectrum has to be sought in a Fock space constructed from the
space of the Coulomb interacting MESs $\{|\alpha )\}$ and the Fock space of
photons $\{|N_{\mathrm{ph}}\rangle\}$
\begin{equation}
      |\alpha )\otimes|N_{\mathrm{ph}}\rangle
      \longrightarrow |\alpha\rangle_{\mathrm{e-EM}}.
\end{equation}
The diagonalization of the electron-photon Hamiltonian (\ref{H-e-EM}) yields
\begin{equation}
      |\breve{\mu}) = \sum_{\alpha}{\cal W}_{\mu\alpha}|\alpha\rangle_{\mathrm{e-EM}} .
\label{W}
\end{equation}

\begin{figure}[htbq]
      \includegraphics[width=0.234\textwidth,angle=0,viewport=16 25 350 668,clip]{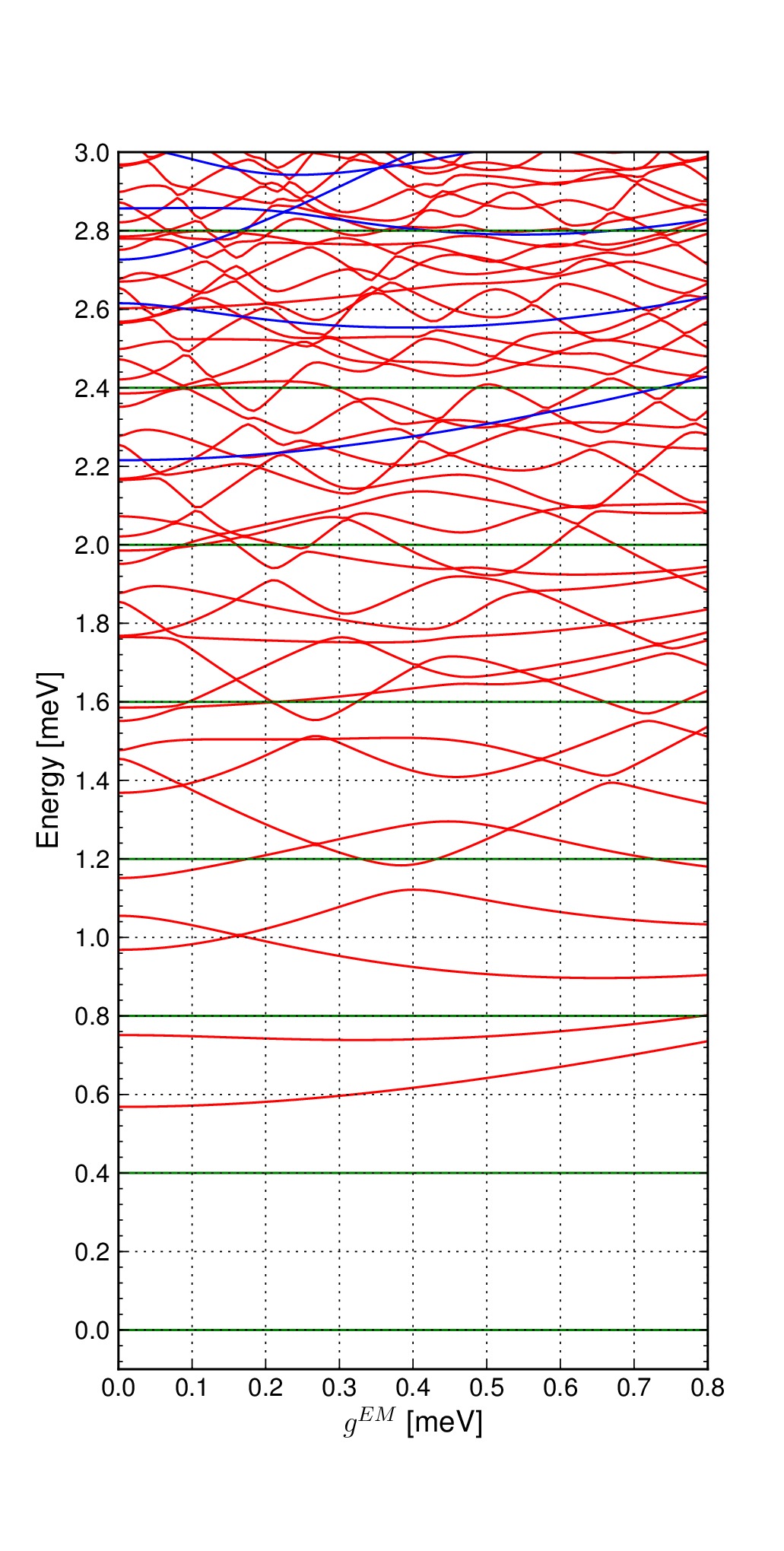}
      \includegraphics[width=0.234\textwidth,angle=0,viewport=16 25 350 668,clip]{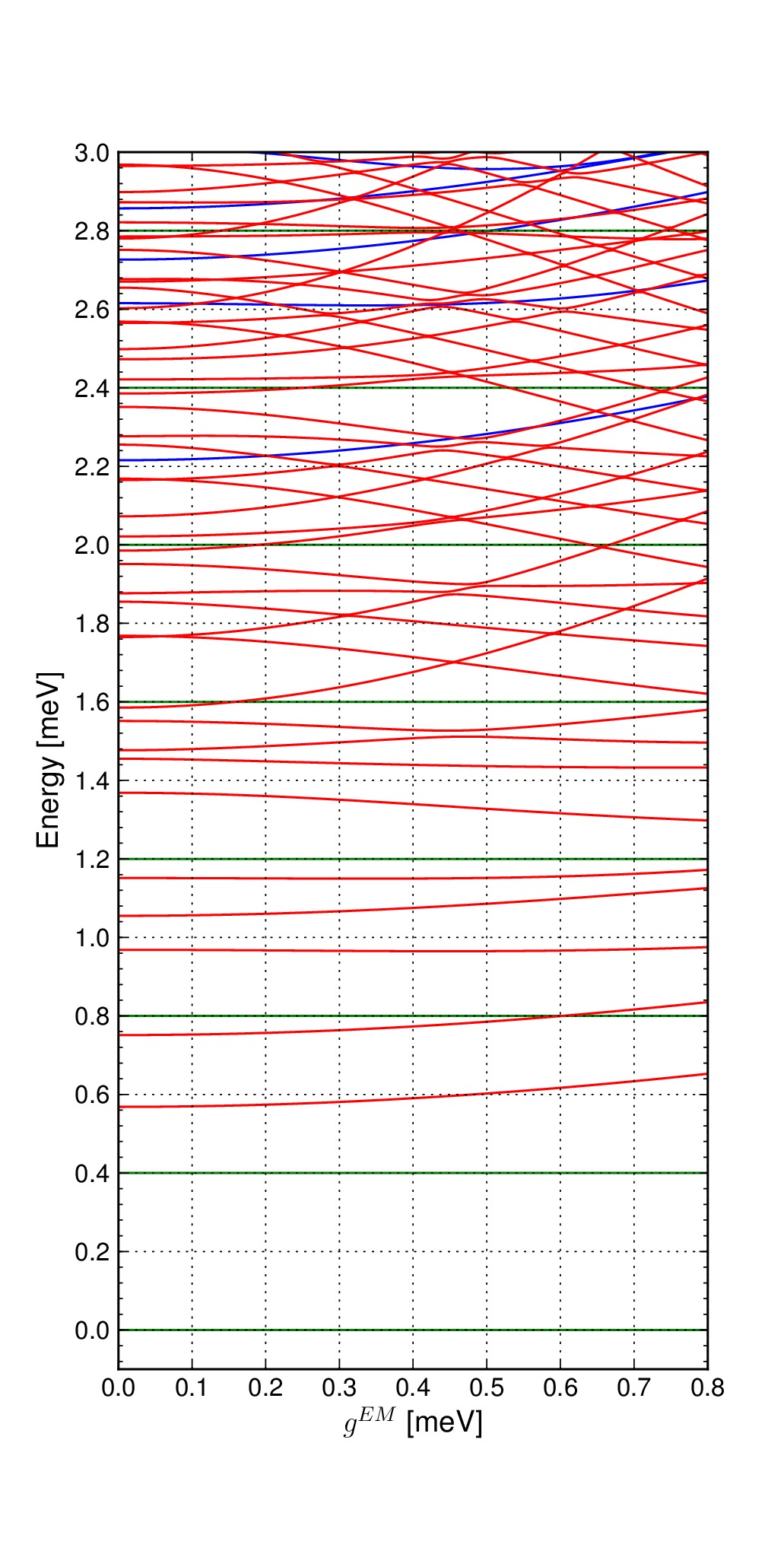}
      \caption{(Color online) The many-body energy spectra for Coulomb interacting electrons
               coupled to quantized cavity photon modes with the electric component polarized
               along the finite quantum wire (x-polarization, left panel), and perpendicular to the
               wire (y-polarization, right panel) versus the electron-photon coupling strength 
               $g^\mathrm{EM}={\cal E}_\mathrm{c}$. In the energy range shown there are 
               states with no electrons (green), one electron (red), and two electrons (blue).
               $B=0.1$ T, $\hbar\Omega_0=1.0$ meV, $\hbar\omega = 0.4$ meV,
               $L_x=300$ nm, $m^*=0.067m_e$, $\kappa =12.4$, the dielectric constant of GaAs.
               }
      \label{E_rof}
\end{figure}

The general scheme here is to start the numerical calculations with a fairly large
number of SESs, $N_\mathrm{SES}$, and retain only a certain number of MESs in the
energy range that is relevant to the transport later, because the total number of MESs,
$N_\mathrm{MES}$, may be too large for the subsequent calculations. 
We will call a further twist on this procedure: ``A stepwise introduction of complexity
to the model and a stepwise truncation of its many-body space''. First, we introduce the
Coulomb interaction between the electrons, second, we truncate the huge many-electron space.
Then we add the Hamiltonian of photons and the electron-photon interaction, 
and again undertake a truncation of the
many-body space in order to have a number of Many Body States (MBSs) 
for which the transport calculation can be performed.  
The MBSs are eigenstates of the interacting electron-photon system.
We stress here that the unitary transformations necessary between the
different many-body spaces, Eqs (\ref{V}) and (\ref{W}), have to be completed 
before truncation. In actual numbers, in the transport calculations here 
for $g^\mathrm{EM}={\cal E}_\mathrm{c}\leq 0.1$ meV we used 10 SESs resulting in
1024 MBSs. The Fock-space of Coulomb interacting electrons was then truncated to the 
lowest 64 states. The inclusion of the single photon mode was accomplished with 27
photon states and the ensuing 1728 photon-electron many-body states will be truncated to
the 64 states lowest in energy before the onset of the transport calculation.
In order to present the energy spectra for the closed electron-photon system
in Fig.\ \ref{E_rof} we use 200 electron states and 20 photon states
for the much larger values of the coupling $g^\mathrm{EM}$. For the closed
system we calculate the energy-spectra for each number of electrons present
separately. We present the calculations for the closed system in a separate 
publication with more details on the convergence of the calculations and
properties of the system for high values of the coupling constant.
\cite{Jonasson2011:01} There we also discuss when the second order term 
in $\mathbf{A}$ in the electron-photon interaction is necessary for our
model of a finite quantum wire.

The many-body energy spectrum of the electron-photon states are shown in Fig.\ \ref{E_rof} for the
two polarizations, along the finite quantum wire (x-polarization) and perpendicular to it (y-polarization).
The horizontal states (green) in Fig.\ \ref{E_rof} are states only with photons and no electrons
that are thus independent of the coupling of the electrons and the photons. 

\begin{figure}[htbq]
      \includegraphics[width=0.42\textwidth,angle=0,viewport=1 1 338 244,clip]{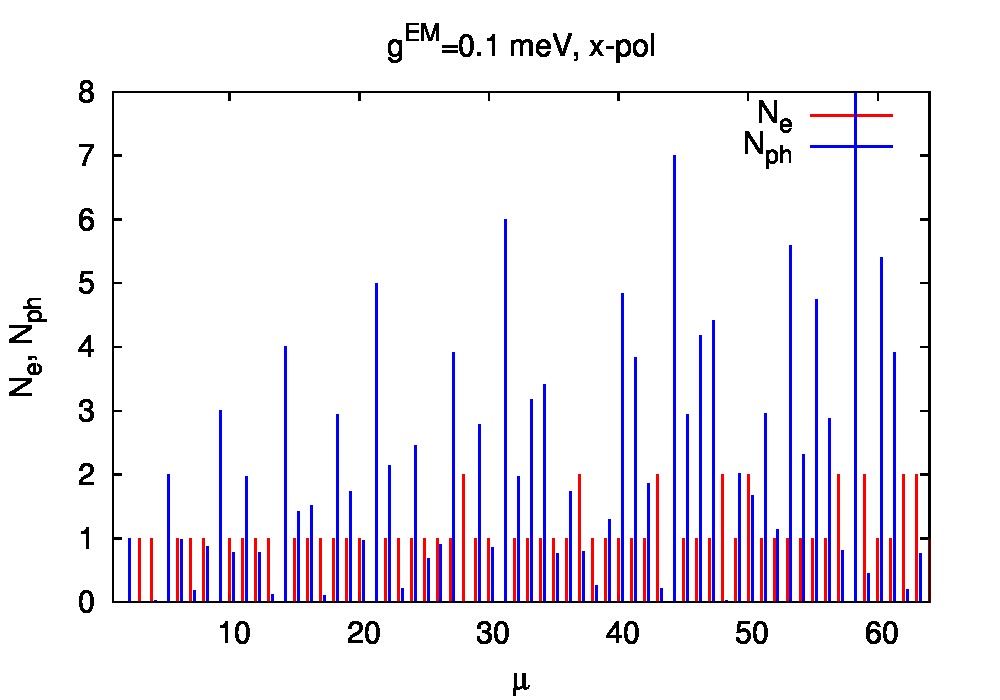}
      \caption{(Color online) The number of electrons in the many-body state $|\breve{\mu})$ (red), and the
               diagonal element of the photon-number operator, $N^\mathrm{ph}_{\mu\mu}$ (blue).
               $B=0.1$ T, $g^\mathrm{EM}=0.1$ meV, $\hbar\Omega_0=1.0$ meV, $\hbar\omega = 0.4$ meV,
               $L_x=300$ nm, $m^*=0.067m_e$, $\kappa =12.4$.}
      \label{N-virkjar}
\end{figure}

In order to gain further insight into the character of the many-body states in Fig.\ \ref{E_rof}
we plot in Fig.\ \ref{N-virkjar} the number of electrons in each state $|\breve{\mu})$, a conserved
quantity, and thus an integer. In the same plot we add the diagonal of the photon number operator
$N_\mathrm{ph}=a^\dagger a$ in the appropriate space $\{|\breve{\mu})\}$. The photon number is not
a conserved quantity in the closed system, but the diagonal of the operator should give us some
indication of the strong mixing of the photon-electron states by the cavity coupling.
Indeed, we see a considerable photon content in many states, not only the few ones that 
contain no electrons (i.e.\ $\mu = 1,2,5,9,14,21,31,44,58$). 

The spectra versus the coupling constant $g^\mathrm{EM}$ in Fig.\ \ref{E_rof} have resemblance with 
the energy spectrum of a single electron versus the magnetic field $B$ in a non-circular 
quantum dot.\cite{Ingibjorg99:16591,Magnusdottir00:10229} The reason comes clear when the
single-electron Hamiltonian for the electron is written in terms of lowering and raising operators
and compared to the many-body e-EM Hamiltonian (\ref{H-e-EM}). As long as the $A^2$-term is included
the magnetic field in the dot, $B$, and the couling to the photons,
$g^\mathrm{EM}$, in our model play a similar role. The $xy$-symmetry breaking is not only 
caused by the geometry of the central system here, but also by the polarization of the photon field.

Here, we have coupled a many-level electron system with photons
using the full electromagnetic coupling (\ref{H-e-VEMV}), and it is interesting to compare the
results with spectra obtained for the Jaynes-Cummings model without the rotating wave 
approximation derived by Feranchuk et al.\cite{Feranchuk96:4035} and Li et al.\cite{Li09:044212} 
This comparison will be detailed in another publication concentrated on the properties of the
closed system.\cite{Jonasson2011:01}

\subsection{System connected to leads}
The closed system of Coulomb interacting electrons interacting with the 
single electromagnetic cavity mode is coupled at $t=0$ to external leads
acting as electron reservoirs. We use a formalism proposed by 
Nakajima and Zwanzig to project the time evolution of the system onto the
central system by partial tracing operations with respect to the operators
of the leads.\cite{Nakajima58:948,Zwanzig60:1338} The coupling Hamiltonian is
of the form
\begin{equation}
      H_\mathrm{T}(t)=\sum_{i,l}\chi^l(t)\int dq\;
      \left\{T^l_{qi}c_{ql}^\dagger d_i + (T^l_{qa})^*d_i^\dagger c_{ql} \right\},
\label{H_T}
\end{equation}
where $l=\{ L,R\}$ referring to the left and the right lead, and $\chi^l(t)$ is
the time-dependent switching function of the coupling. The semi-infinite leads
are in the same perpendicular constant external magnetic field $\mathbf{B}$ as
the central system. Their energy spectra are continuous bands and the integral in Eq.\ (\ref{H_T})
represents a summation over the band index and an integral over the continuous
``momentum'' $q$ from the band bottom to an appropriate band cut-off.
The coupling tensor $T^l_{qi}$ of a single-electron states
$|q\rangle$ in the lead $l$ to states $|i\rangle$ in the system is modeled as a non-local
overlap integral of the corresponding wave functions in the contact
regions of the system, $\Omega_S^l$, and the lead $l$,
$\Omega_l$\cite{Gudmundsson09:113007}
\begin{equation}
      T^l_{iq} = \int_{\Omega_S^l\times \Omega_l} d{\bf r}d{\bf r}'
      \left[\psi^l_q ({\bf r}') \right]^*\psi^S_i({\bf r})\;
      g^l_{iq} ({\bf r},{\bf r'}).
\label{T_aq}
\end{equation}
The function
\begin{align}
      g^l_{iq} ({\bf r},{\bf r'}) =
                   g_0^l&\exp{\left[-\delta_1^l(x-x')^2-\delta_2^l(y-y')^2\right]}\nonumber\\
                   &\times\exp{\left(\frac{-|E_i-\epsilon^l(q)|}{\Delta_E^l}\right)}.
\label{gl}
\end{align}
with ${\bf r}\in\Omega_\mathrm{S}^l$ and ${\bf r}'\in\Omega_l$
defines the `nonlocal overlap' and their affinity in energy. 

The Liouville-von Neumann equation describing the time-evolution of the total
system, the finite quantum wire, the cavity photons, and the leads
\begin{equation}
      i\hbar\dot W(t)=[H(t),W(t)],\quad W(t<0)=\rho_\mathrm{L}\rho_\mathrm{R}\rho_\mathrm{S},
\label{L-vN}
\end{equation}
where $W$ is the statistical operator of the total system
and the equilibrium density operator of the disconnected lead $l\in\{L,R\}$ with
chemical potential $\mu_l$ is
\begin{equation}
      \rho_l=\frac{e^{-\beta (H_l-\mu_l N_l)}}{{\rm Tr}_l \{e^{-\beta(H_l-\mu_l N_l)}\}}.
\label{rho_l}
\end{equation}

\begin{figure}[htbq]
      \includegraphics[width=0.42\textwidth,angle=0,viewport=1 1 338 240,clip]{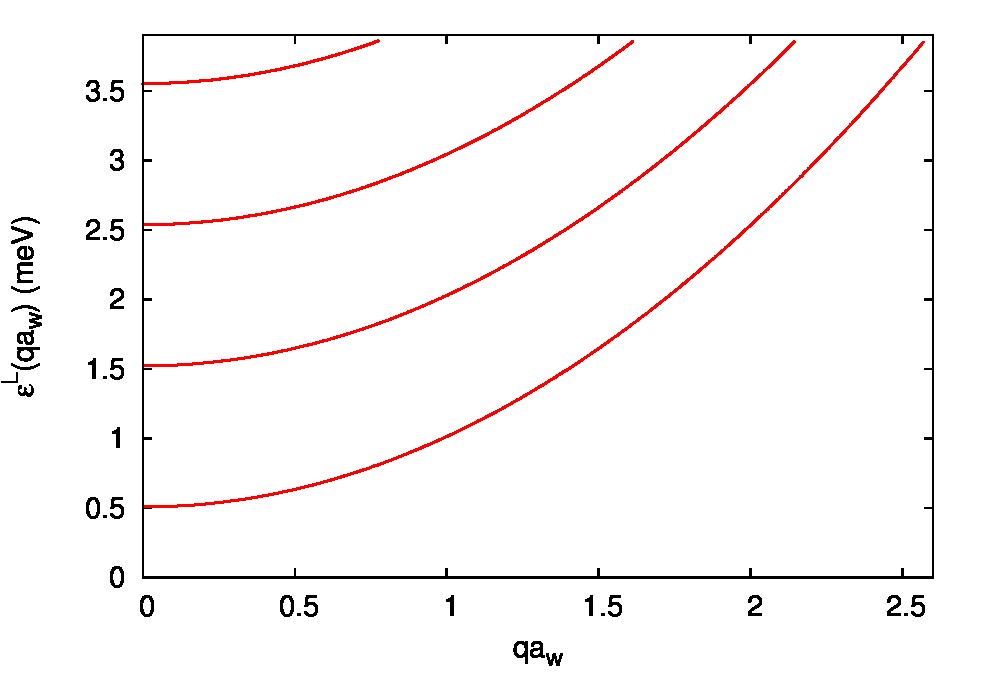}
      \caption{The single electron energy spectrum of the leads.
               $B=0.1$ T, $\hbar\Omega_0^l=1.0$ meV.}
      \label{Eql}
\end{figure}

\begin{figure}[htbq]
      \includegraphics[width=0.20\textwidth,angle=0,viewport=1 1 233 115,clip]{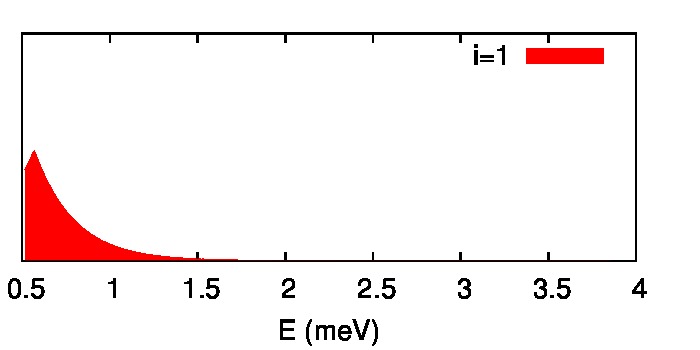}
      \includegraphics[width=0.20\textwidth,angle=0,viewport=1 1 208 100,clip]{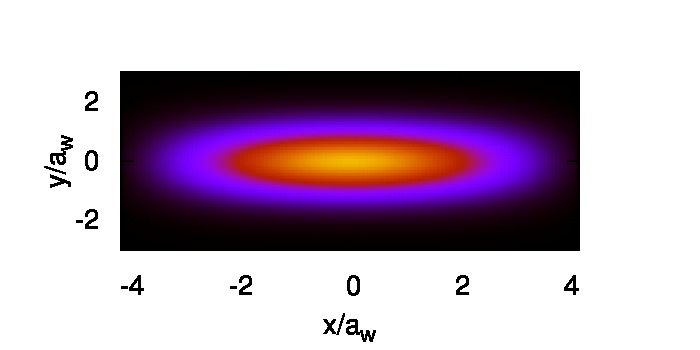}
      \includegraphics[width=0.20\textwidth,angle=0,viewport=1 1 233 115,clip]{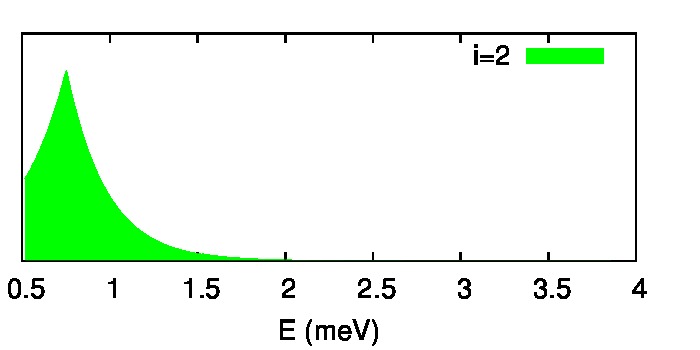}
      \includegraphics[width=0.20\textwidth,angle=0,viewport=1 1 208 100,clip]{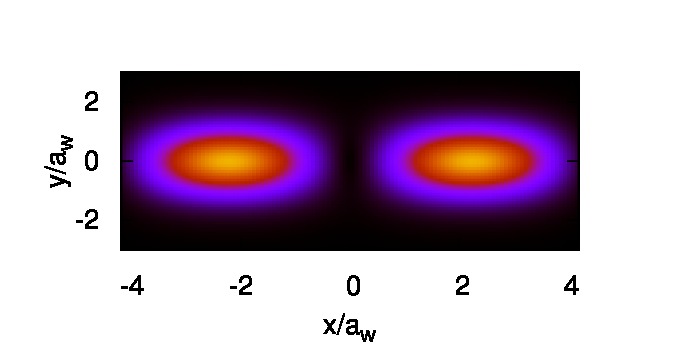}
      \includegraphics[width=0.20\textwidth,angle=0,viewport=1 1 233 115,clip]{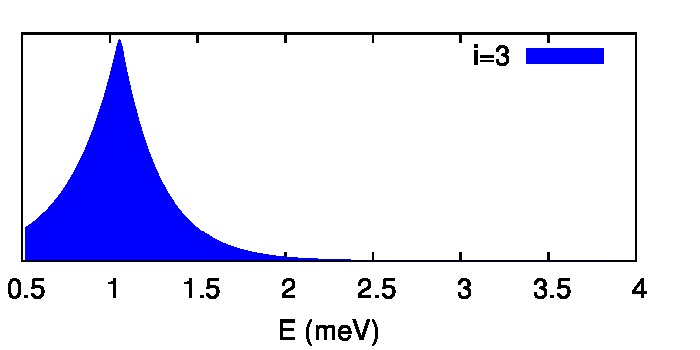}
      \includegraphics[width=0.20\textwidth,angle=0,viewport=1 1 208 100,clip]{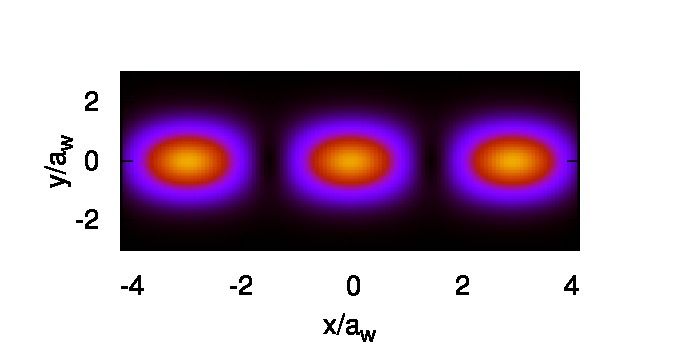}
      \includegraphics[width=0.20\textwidth,angle=0,viewport=1 1 233 115,clip]{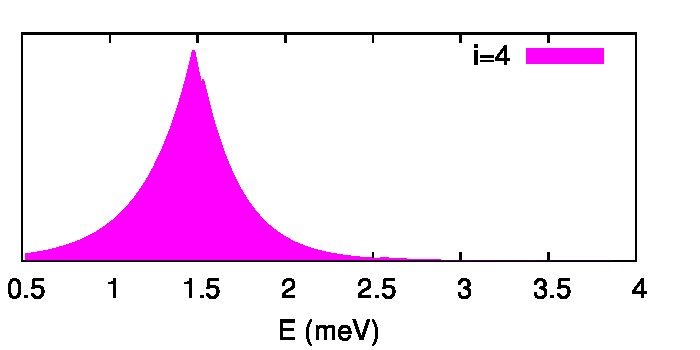}
      \includegraphics[width=0.20\textwidth,angle=0,viewport=1 1 208 100,clip]{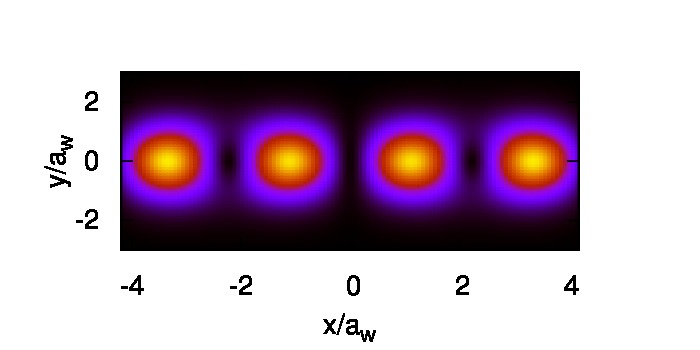}
      \includegraphics[width=0.20\textwidth,angle=0,viewport=1 1 233 115,clip]{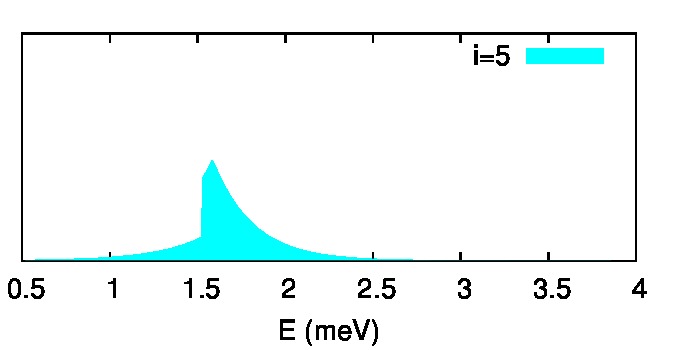}
      \includegraphics[width=0.20\textwidth,angle=0,viewport=1 1 208 100,clip]{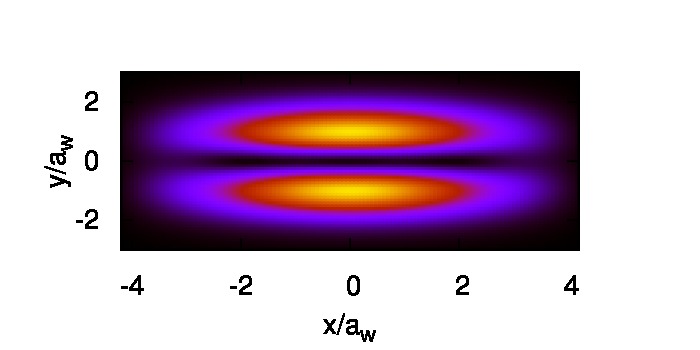}
      \includegraphics[width=0.20\textwidth,angle=0,viewport=1 1 233 115,clip]{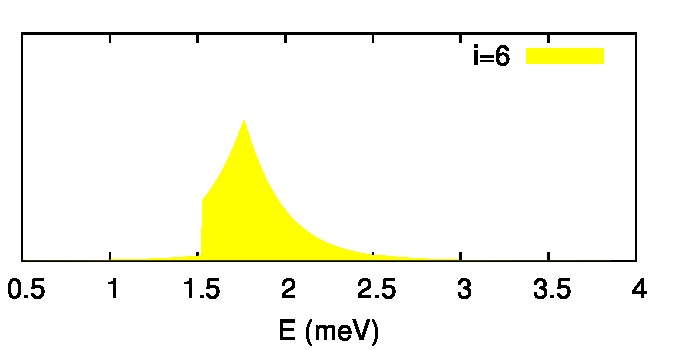}
      \includegraphics[width=0.20\textwidth,angle=0,viewport=1 1 208 100,clip]{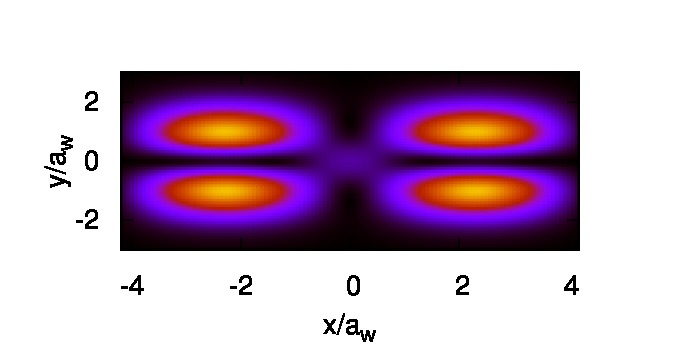}
      \includegraphics[width=0.20\textwidth,angle=0,viewport=1 1 233 115,clip]{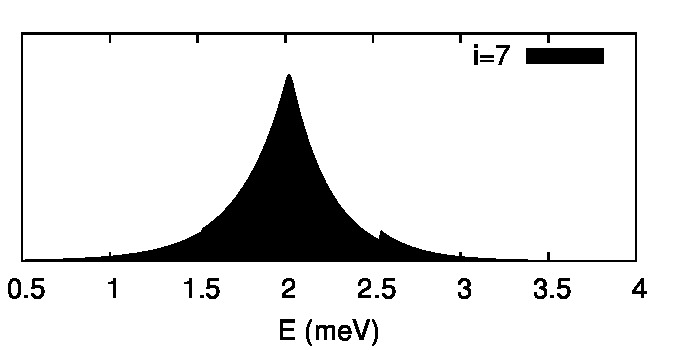}
      \includegraphics[width=0.20\textwidth,angle=0,viewport=1 1 208 100,clip]{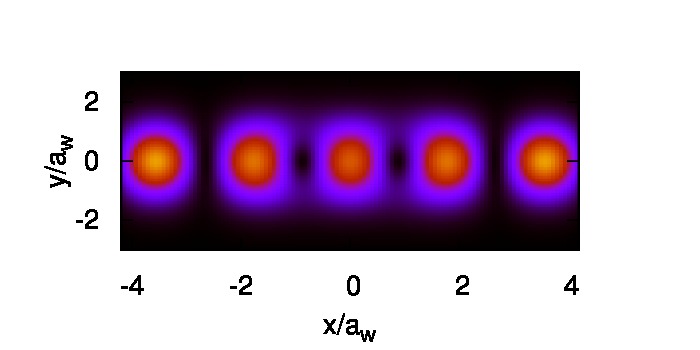}
      \includegraphics[width=0.20\textwidth,angle=0,viewport=1 1 233 115,clip]{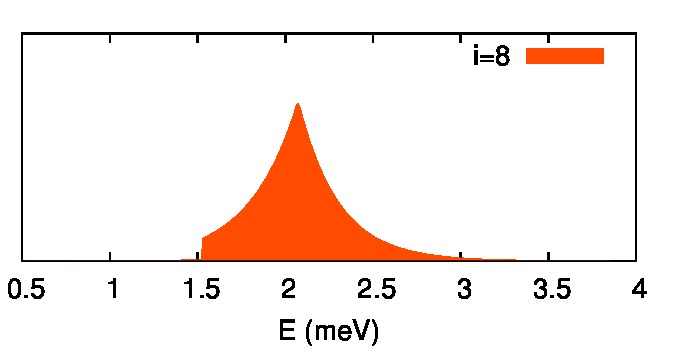}
      \includegraphics[width=0.20\textwidth,angle=0,viewport=1 1 208 100,clip]{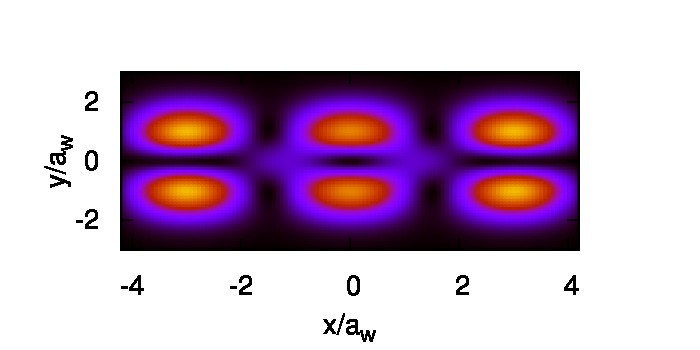}
      \includegraphics[width=0.20\textwidth,angle=0,viewport=1 1 233 115,clip]{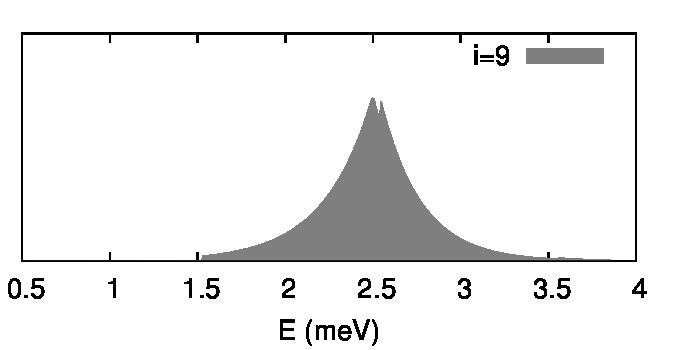}
      \includegraphics[width=0.20\textwidth,angle=0,viewport=1 1 208 100,clip]{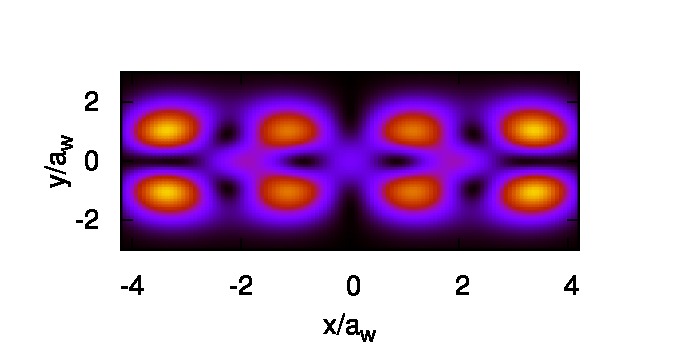}
      \includegraphics[width=0.20\textwidth,angle=0,viewport=1 1 233 115,clip]{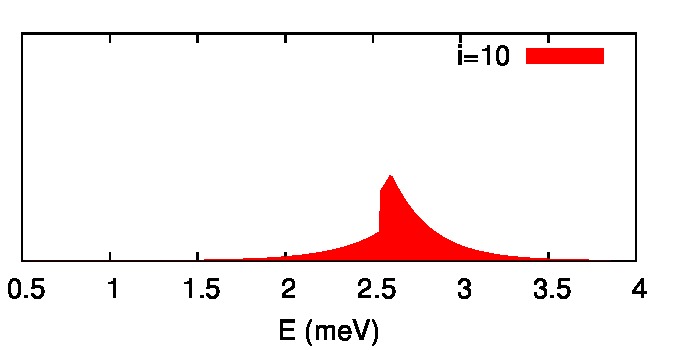}
      \includegraphics[width=0.20\textwidth,angle=0,viewport=1 1 208 100,clip]{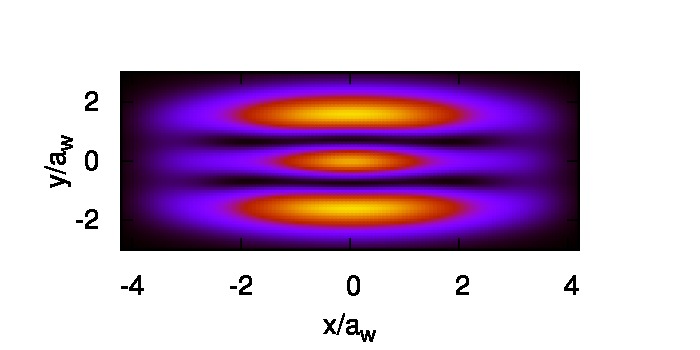}
      \caption{(Color online) The spectral density $J_i^l(E)$ (left) and the probability amplitude for the
               corresponding single-electron eigenstate (right). $B=0.1$T, $\hbar\Omega_0=1.0$ meV,
               $L_x=300$ nm, $\Delta^l_E=0.25$ meV, $g_0^la_w^{3/2}=53.14$ meV,
               $\delta_{1,2}^la_w^2=0.4916$.}
      \label{JE}
\end{figure}

Commonly, the spectral density $J_a(\omega )$ is used to describe the coupling of states
in an open system to reservoirs or leads.\cite{Chen11:115439,Jin11:053704}
We do not use it explicitly here, but the spectral density for lead $l$ with respect
to the SES $a$ in the central system
\begin{equation}
      J_i^l(E) \propto \int dq |T^l_{iq}|^2 \delta (E - \epsilon^l(q))
\label{Jla}
\end{equation}
is a convenient tool to demonstrate graphically the phenomenological coupling selected here,
in Equations (\ref{T_aq}) and (\ref{gl}). In order to do this we show first in Fig.\ \ref{Eql}
the single-electron energy spectrum in the leads 
and in Fig.\ \ref{JE} the spectral density $J_i^l(E)$ and the probability density
for each SES used to build the MESs of the system.   
We see in Fig.\ \ref{JE} from jumps in the spectral density that SESs belonging to different 
subbands in the central system have different coupling strengths to the subbands in the leads as 
a result of their symmetry. The simple geometry of the central system here does not result in
states with widely different localization character. 

The Liouville-von Neumann equation
is projected on the central system of electrons and photons by partial tracing operations with respect to
the operators of the leads. Defining the reduced density operator (RDO) of the central
system
\begin{equation}
      \rho_\mathrm{S}(t)={\rm Tr}_\mathrm{L} {\rm Tr}_\mathrm{R} W(t),
      \quad \rho_\mathrm{S}(0)=\rho_\mathrm{S},
\label{rdo}
\end{equation}
we obtain an integro-differential equation for the RDO, the generalized
master equation (GME)
\begin{align}
      &\dot{\rho}_\mathrm{S}(t)=\frac{i}{\hbar}\left[H_\mathrm{S}, {\rho}_\mathrm{S}(t)\right]\nonumber\\
      &-\frac{1}{\hbar^2}{\mathrm{Tr}_\mathrm{res}}\left\{\left[ H_\mathrm{T}(t'),
      \int_0^t dt'\; \left[ U(t-t')H_\mathrm{T}(t')U^+(t-t')\right.\right.\right. \nonumber\\
      &\quad\quad\quad \quad\quad\quad
      ,\left.\left.\left. U_0(t-t')\rho_\mathrm{S}(t')U_0^+(t-t')\right]
      \right]\right\},
\label{RDO}
\end{align}
with the time evolution operator for the isolated systems of Coulomb interacting electrons 
interacting with photons, and noninteracting electrons in the leads,
$U(t)=\exp{\{-i(H_\mathrm{e}+H_\mathrm{Coul}+H_\mathrm{EM}+H_L+H_R)t/\hbar\}}$, 
without the coupling $H_\mathrm{T}(t)$.
The time evolution of the closed isolated system of Coulomb interacting electrons 
interacting with the photons is governed by $U_0(t)=\exp{\{-iH_0t/\hbar\}}$.

The equation for the RDO (\ref{RDO}) can be cast into a coupled
set of integro-differential equations for the matrix elements of the RDO by projecting
it on the $\{ |\breve{\mu})\}$-basis in a similar manner as was done for the 
Coulomb interacting electrons in our former work without 
photons.\cite{Moldoveanu10:155442,Gudmundsson10:205319}
The kernel of the equation of motion for the RDO (\ref{RDO}) has been approximated 
to the second order for the lead-system coupling $H_\mathrm{T}(t)$, but due to the integral
structure of the equation the coupling is present in the solution to higher order.

It is convenient to express many operators of the system in a particular basis,
i.\ e.\ like the basis $\{|\mu\rangle\}$ of independent electrons, or in the Coulomb interacting
basis before coupling to the photons $\{|\mu )\}$. The time-dependent mean values of these
operators are best calculated by employing a unitary transformation
between the bases. The photon number operator $N_\mathrm{ph}=a^\dagger a$ is best constructed
in the $\{|\mu\rangle_\mathrm{e-EM}\}$-basis, and the unitary transform introduced in Eq.\ (\ref{W})
to the $\{|\breve{\mu} )\}$-basis yields
\begin{align}
      \langle a^\dagger a\rangle_\mathrm{e-EM}
      &=\sum_\mu (\breve{\mu}|\rho (t)
      a^\dagger a|\breve{\mu}) \nonumber\\
      &= \mathrm{Tr}_\mathrm{S}\{ \rho (t)
      {\cal W}^+N_\mathrm{ph}{\cal W}\}.
\end{align}
Just as a reminder, the unitary transform is performed before the basis is truncated to the size
of the matrix representing the RDO in the $\{|\breve{\mu} )\}$-basis dictated by the computational
effort in solving the GME (\ref{RDO}).  

The time-dependent average current $J_\mathrm{T}^\mathrm{L}$ into the system from the
left lead, and $J_\mathrm{T}^\mathrm{R}$ from the system into the right lead are
calculated from the GME (\ref{RDO}) along directions introduced in earlier 
publications.\cite{Moldoveanu09:073019,Gudmundsson09:113007,Moldoveanu10:155442}

\section{Transport properties}
For a smooth coupling of the left and the right leads to the central system of electrons
and photons we use a switching function (\ref{T_aq})
\begin{equation}
      \chi^l(t)=\left(1-\frac{2}{e^{\alpha^lt}+1}\right),\quad l\in\{ L,R\}
\label{chi}
\end{equation}
with $\alpha^l = 0.2$ ps$^{-1}$. We fix the temperature of the reservoirs at
$T=0.5$ K, and the coupling strength as $g_0^la_w^{3/2}=53.14$ meV. We select a
bias over the central system by specifying $\mu_L=2.0$ meV, and $\mu_R=1.4$ meV.

\begin{figure}[htbq]
      \includegraphics[width=0.42\textwidth,angle=0,viewport=1 1 346 252,clip]{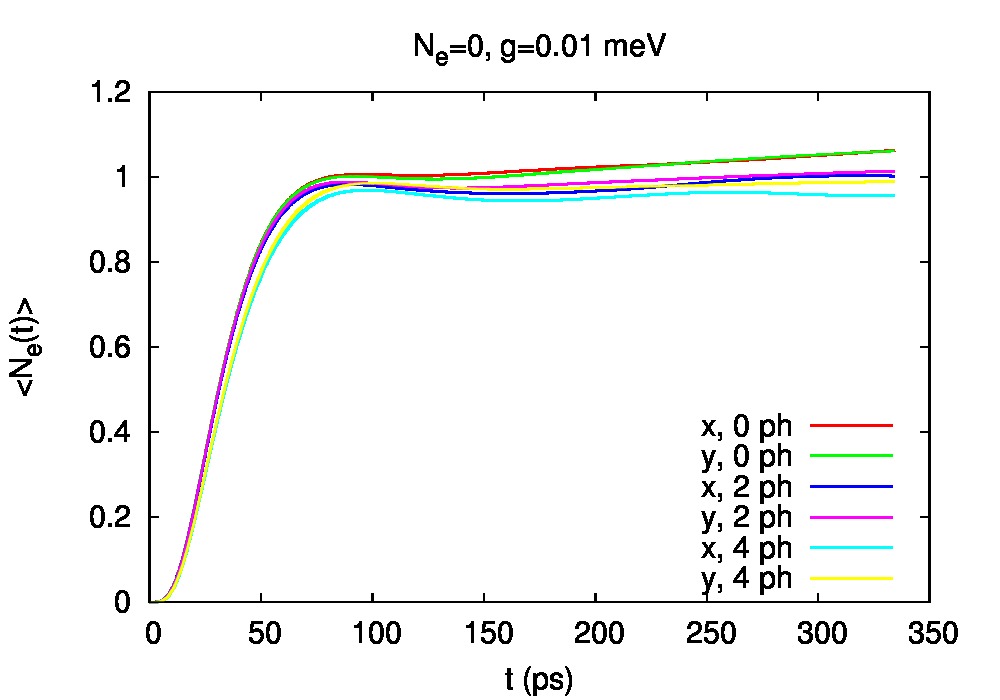}
      \includegraphics[width=0.42\textwidth,angle=0,viewport=1 1 346 252,clip]{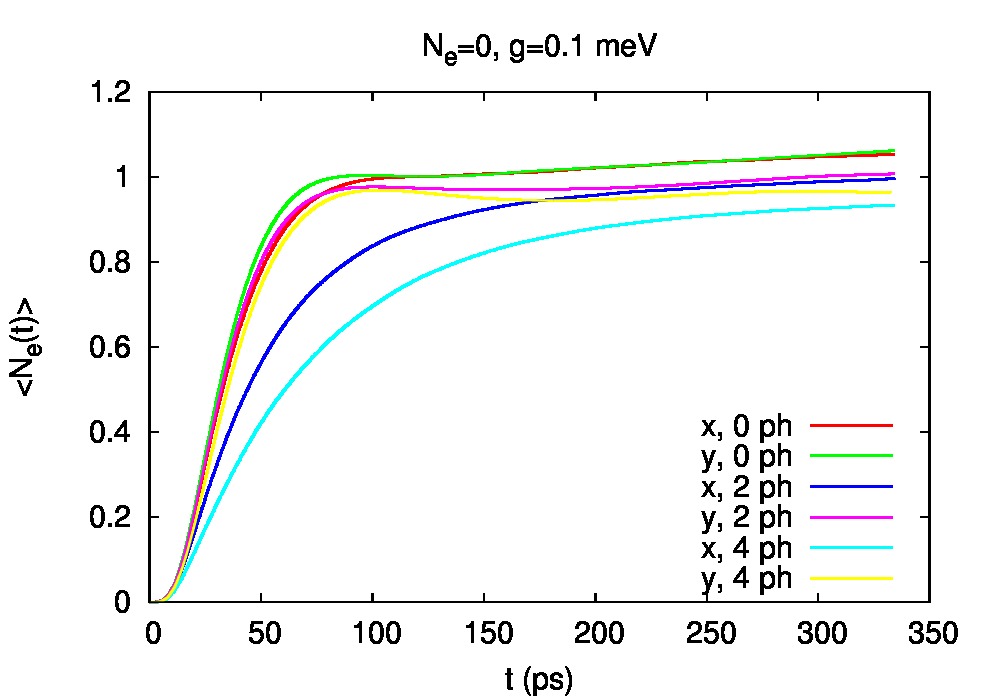}
      \caption{(Color online) The total mean number of electrons $\langle N_\mathrm{e}(t)\rangle$ 
               as a function of time for $g^\mathrm{EM}=0.01$ meV (upper panel) and $g^\mathrm{EM}=0.10$ meV
               (lower panel). $B=0.1$ T, $\hbar\Omega_0=1.0$ meV, $L_x=300$ nm, $\hbar\omega = 0.4$ meV,
               $\mu_L=2.0$ meV, $\mu_R=1.4$ meV, $\Delta^l_E=0.25$ meV, $g_0^la_w^{3/2}=53.14$ meV,
               $\delta_{1,2}^la_w^2=0.4916$, $m^*=0.067m_e$, and $\kappa =12.4$.}
      \label{NetT}
\end{figure}

The time-evolution of the total mean number of electrons $\langle N_\mathrm{e}(t)\rangle$
is displayed in Fig.\ \ref{NetT} for two values of the electron-photon coupling
$g^\mathrm{EM}$. Initially, no electron is present in the central system, but an
integer number of photons is specified with a given polarization.
For the weaker coupling, $g^\mathrm{EM}=0.01$ meV, the charging of the central system seems to
be rather independent of the initial number of cavity photons present and their polarization.
Well after the charging, ($t\sim 300$ ps), a slight difference in the total charge as a function
of the initial condition can be observed, but it is very similar to the difference seen for 
higher coupling, $g^\mathrm{EM}=0.1$ meV. More interestingly, the charging is influenced
by the initial photon number and their polarization for the higher electron-photon coupling.
The charging is slowed down by a higher number of x-polarized photons. The total number of
electrons in the system does not exceed one as could be expected by comparing the values
of the chemical potentials in the leads and the energy spectral in Fig.\ \ref{E_rof}, 
we are in the Coulomb-blocking regime.

For the parameters chosen here describing the geometry and the potentials describing the
central system the aspect ratio of the system for the low energy range is such that states
that can be associated with motion in the $x$-direction are more numerous than states 
describing motion in the $y$-direction as can be verified by a glance at the lowest 10
eigenstates shown in Fig.\ \ref{JE}. We should also have in mind that the photon
energy $\hbar\omega =0.4$ meV $<$ $\hbar\Omega_0=1.0$ meV, the characteristic energy for
parabolic confinement in the $y$-direction. 

\begin{figure}[htbq]
      \includegraphics[width=0.234\textwidth,angle=0,viewport=1 1 330 224,clip]{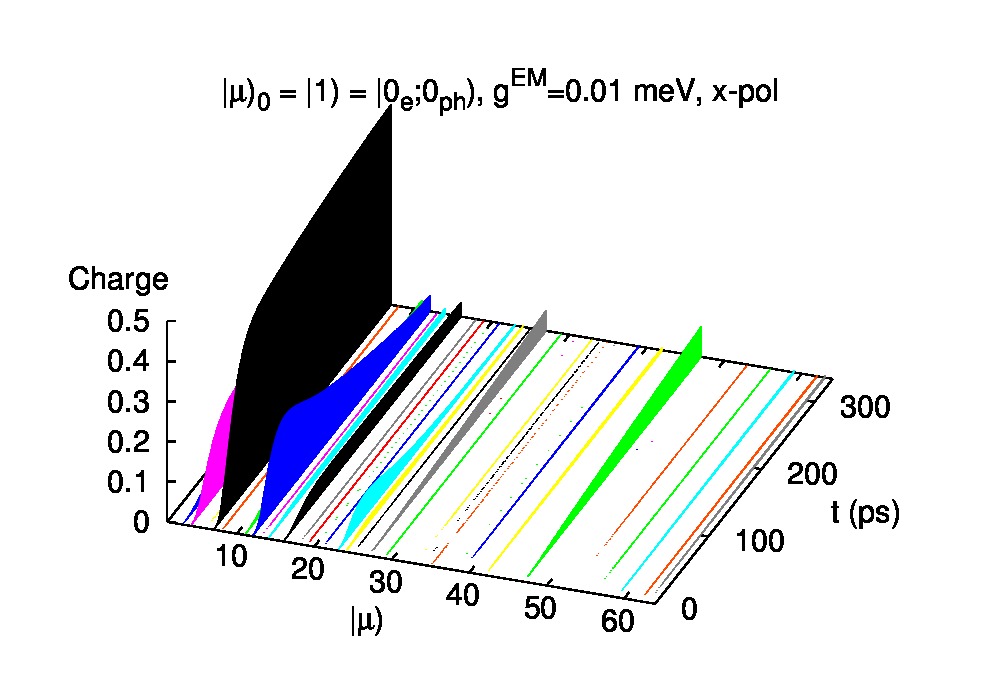}
      \includegraphics[width=0.234\textwidth,angle=0,viewport=1 1 330 224,clip]{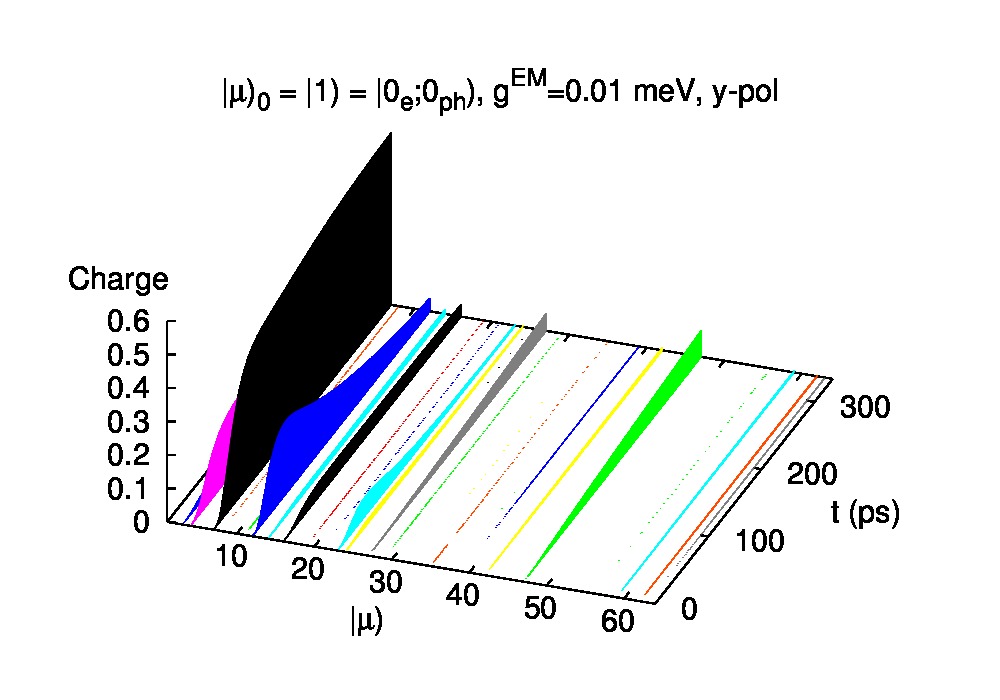}
      \includegraphics[width=0.234\textwidth,angle=0,viewport=1 1 330 224,clip]{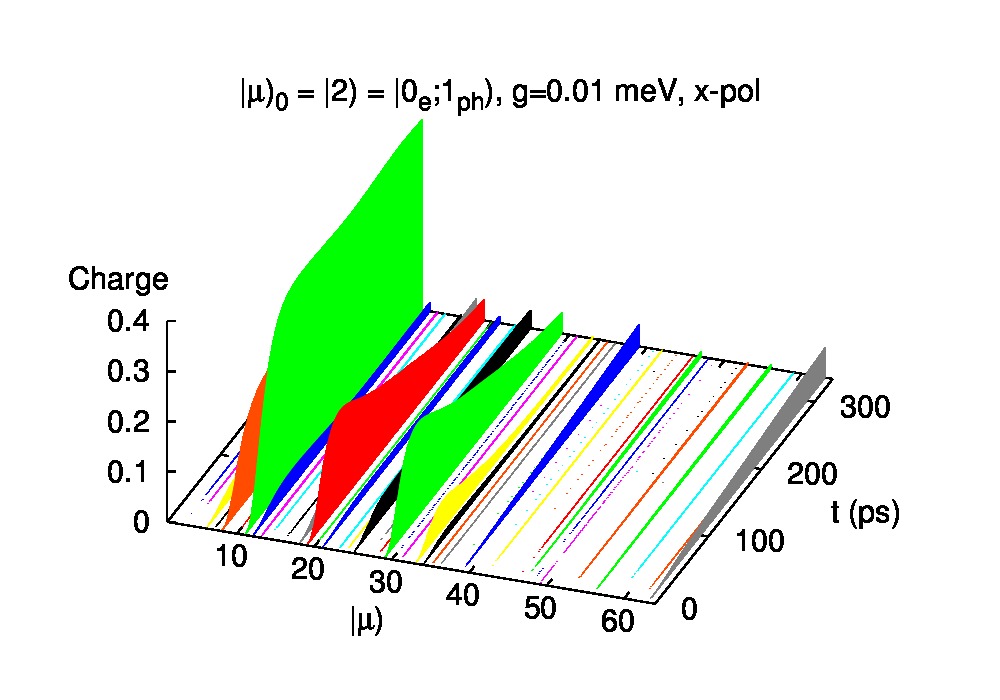}
      \includegraphics[width=0.234\textwidth,angle=0,viewport=1 1 330 224,clip]{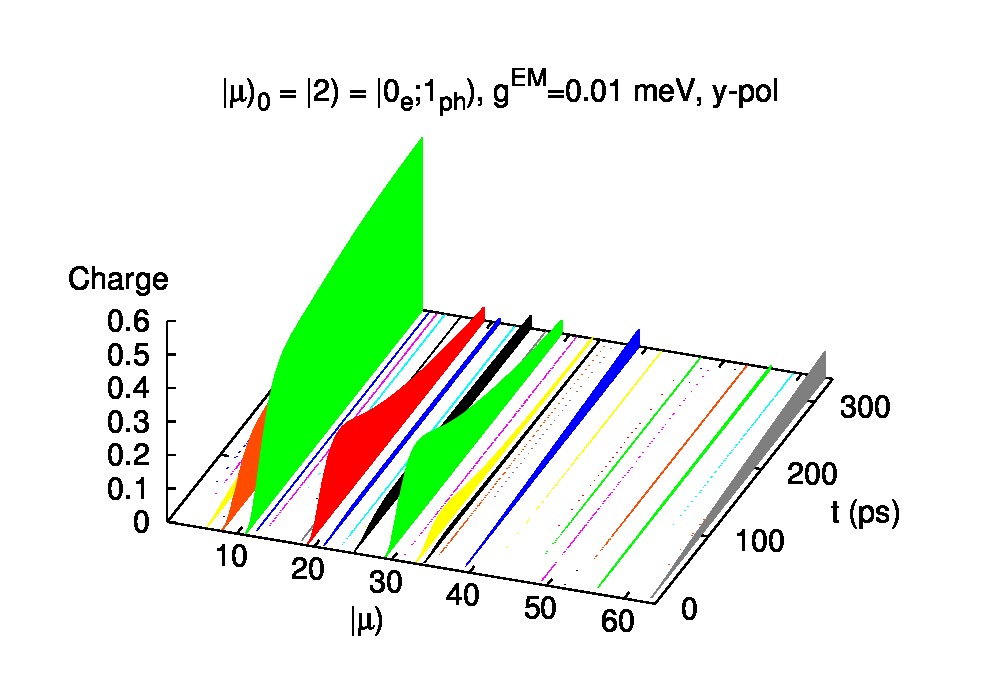}
      \includegraphics[width=0.234\textwidth,angle=0,viewport=1 1 330 224,clip]{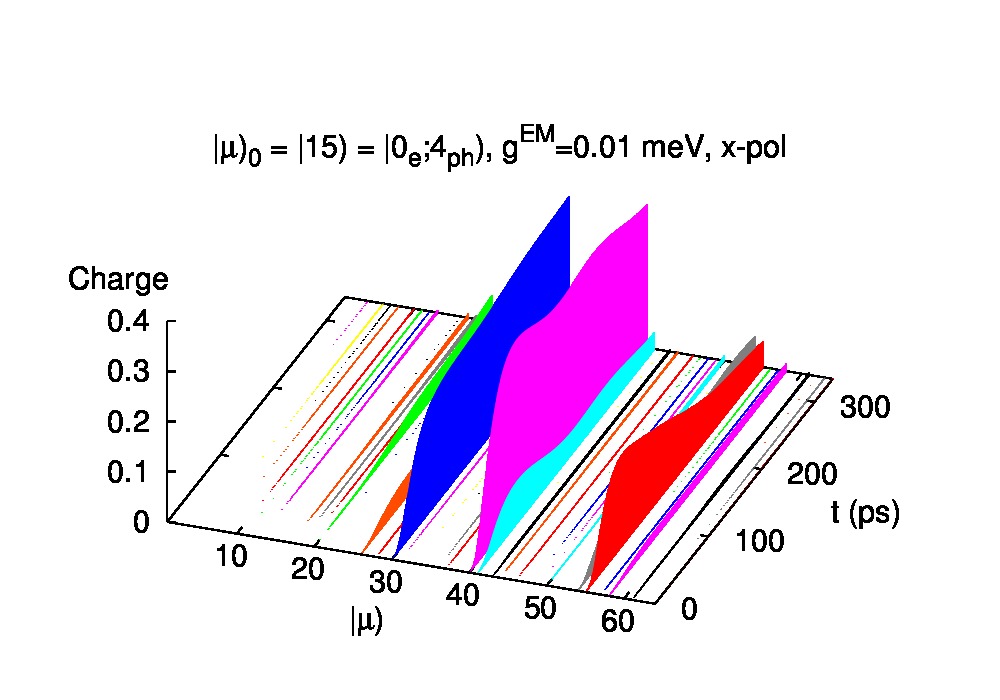}
      \includegraphics[width=0.234\textwidth,angle=0,viewport=1 1 330 224,clip]{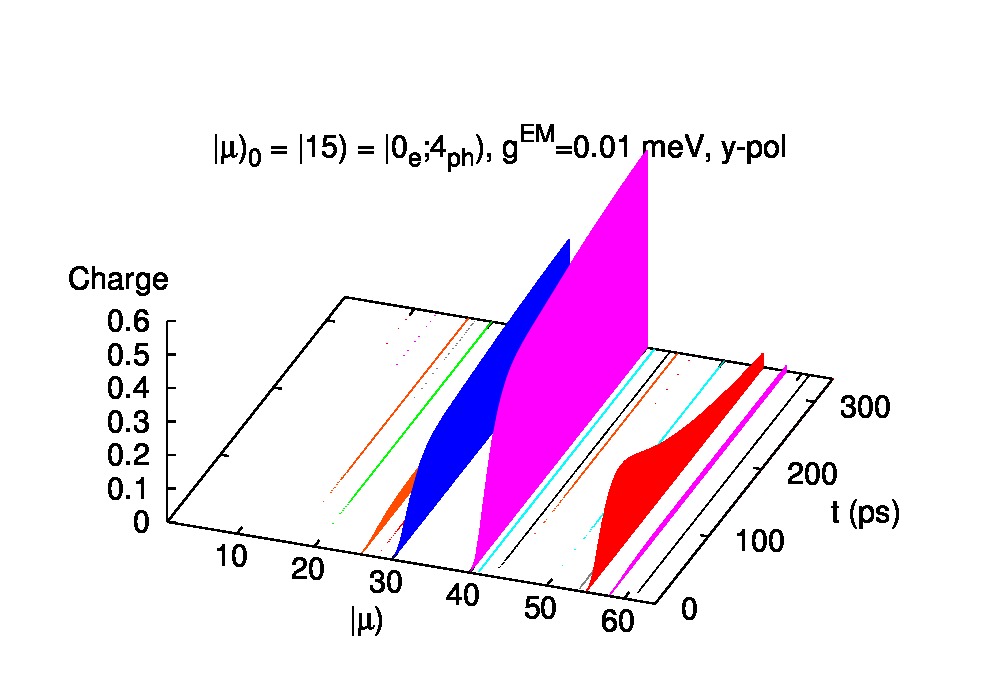}
      \caption{(Color online) The mean number of electrons $\langle N_\mathrm{e}(t)\rangle$ 
               in a MBS $|\breve{\mu})$ for $x$-polarization (left) and $y$-polarization (right)
               as a function of time. Initially, at $t=0$, there is no photon 
               in the cavity (top panels), 1 photon (middle panels), or 4 photons (bottom panels). 
               The initial number of electrons is zero in all cases.
               $B=0.1$ T, $g^\mathrm{EM}=0.01$ meV, $\hbar\omega = 0.4$ meV,
               $\mu_L=2.0$ meV, $\mu_R=1.4$ meV,
               $\hbar\Omega_0=1.0$ meV, $\Delta^l_E=0.25$ meV, $g_0^la_w^{3/2}=53.14$ meV,
               $\delta_{1,2}^la_w^2=0.4916$, $L_x=300$ nm, $m^*=0.067m_e$, and $\kappa =12.4$.}
      \label{Net-g001}
\end{figure}

\begin{figure}[htbq]
      \includegraphics[width=0.234\textwidth,angle=0,viewport=1 1 330 224,clip]{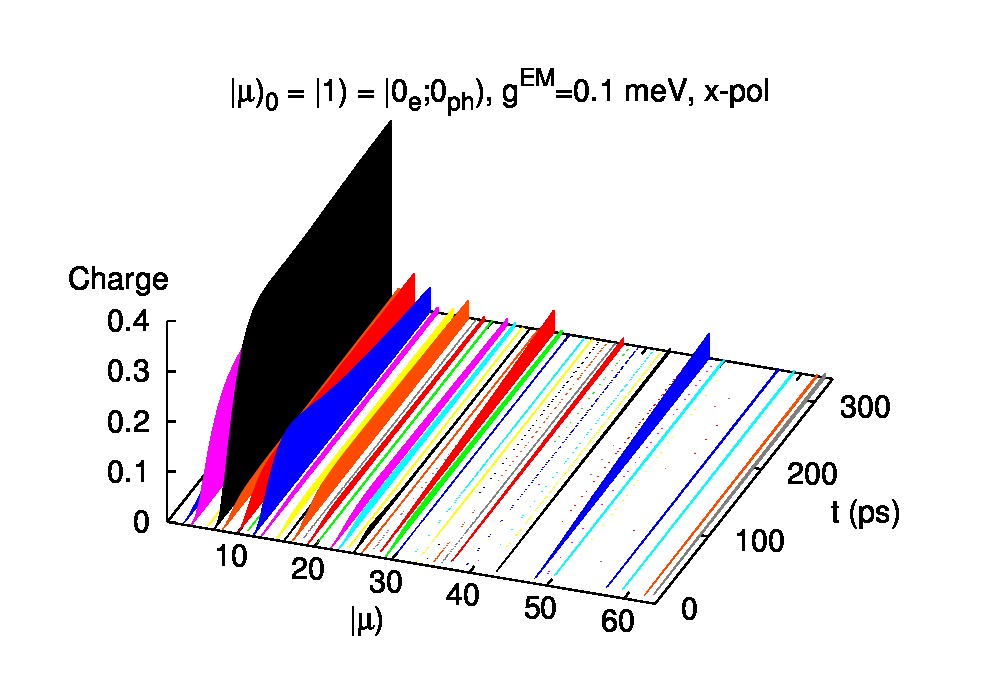}
      \includegraphics[width=0.234\textwidth,angle=0,viewport=1 1 330 224,clip]{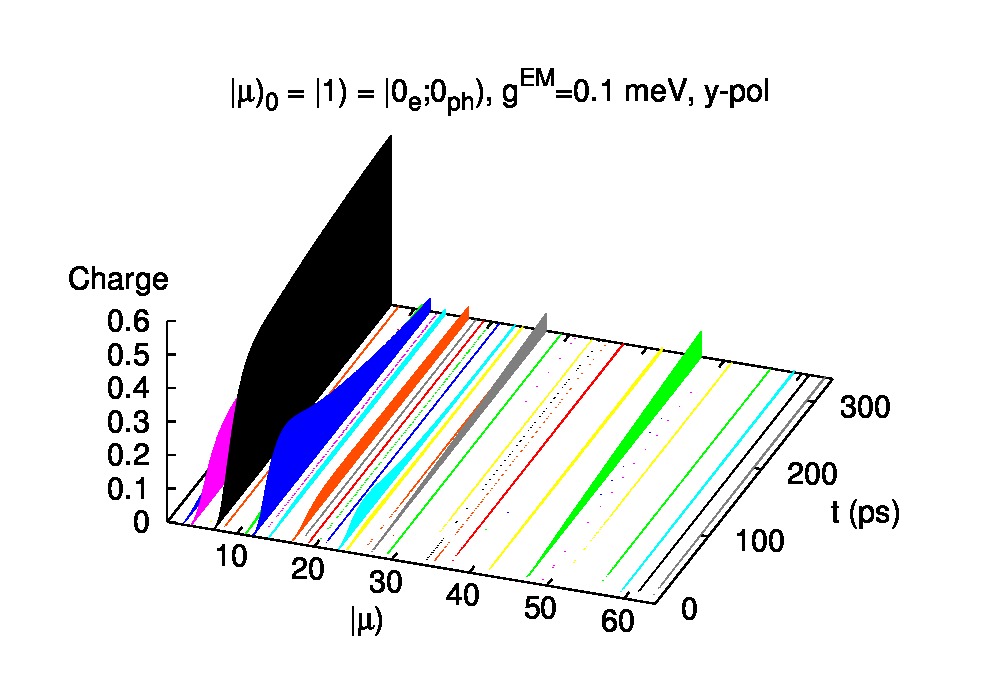}
      \includegraphics[width=0.234\textwidth,angle=0,viewport=1 1 330 224,clip]{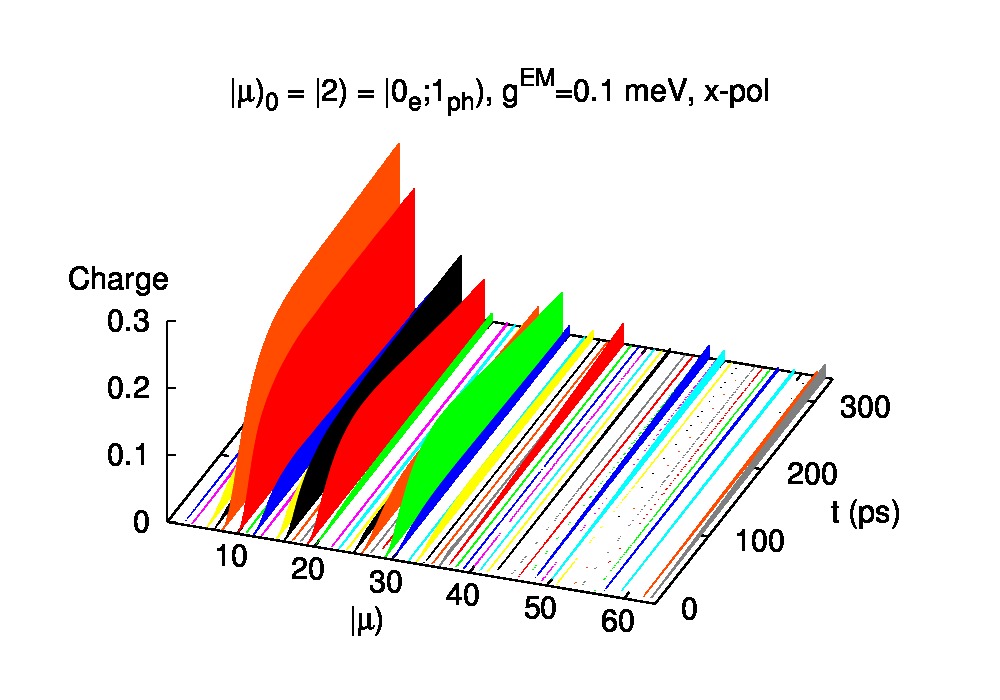}
      \includegraphics[width=0.234\textwidth,angle=0,viewport=1 1 330 224,clip]{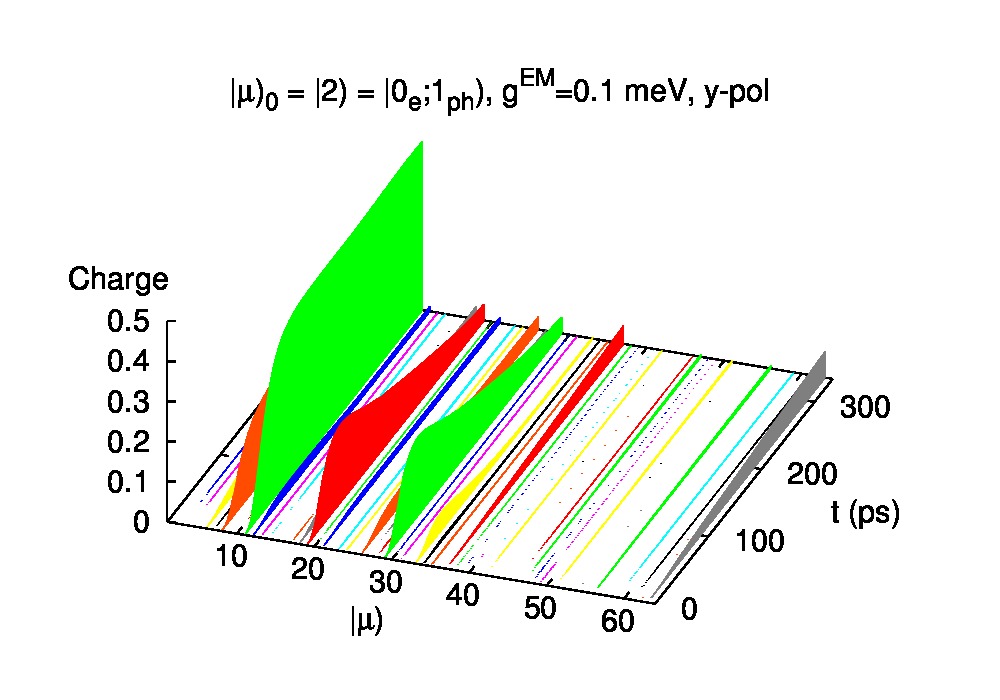}
      \includegraphics[width=0.234\textwidth,angle=0,viewport=1 1 330 224,clip]{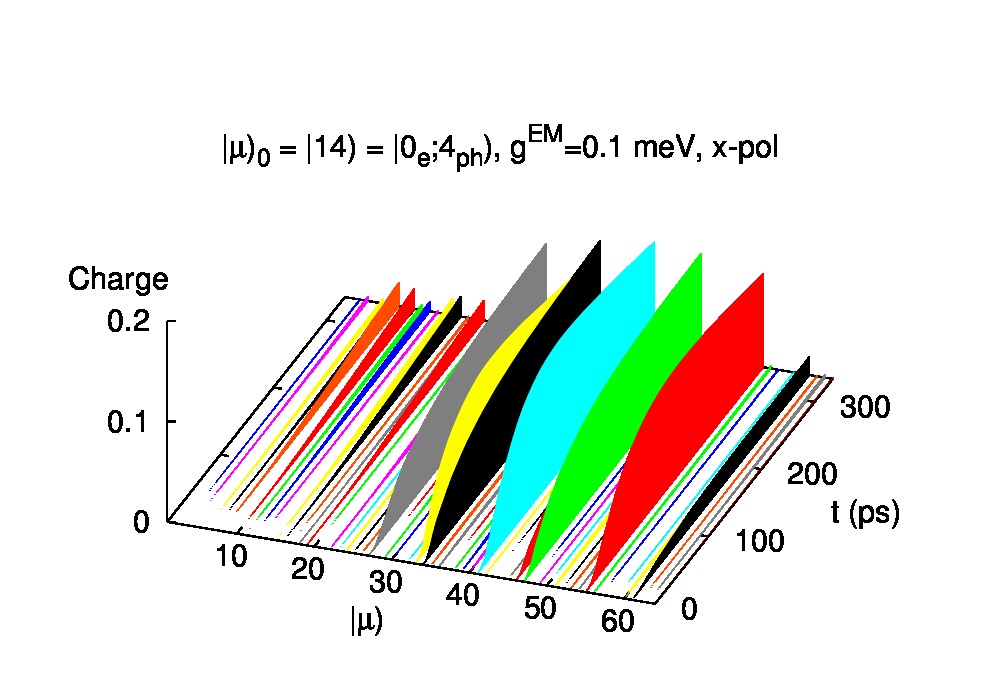}
      \includegraphics[width=0.234\textwidth,angle=0,viewport=1 1 330 224,clip]{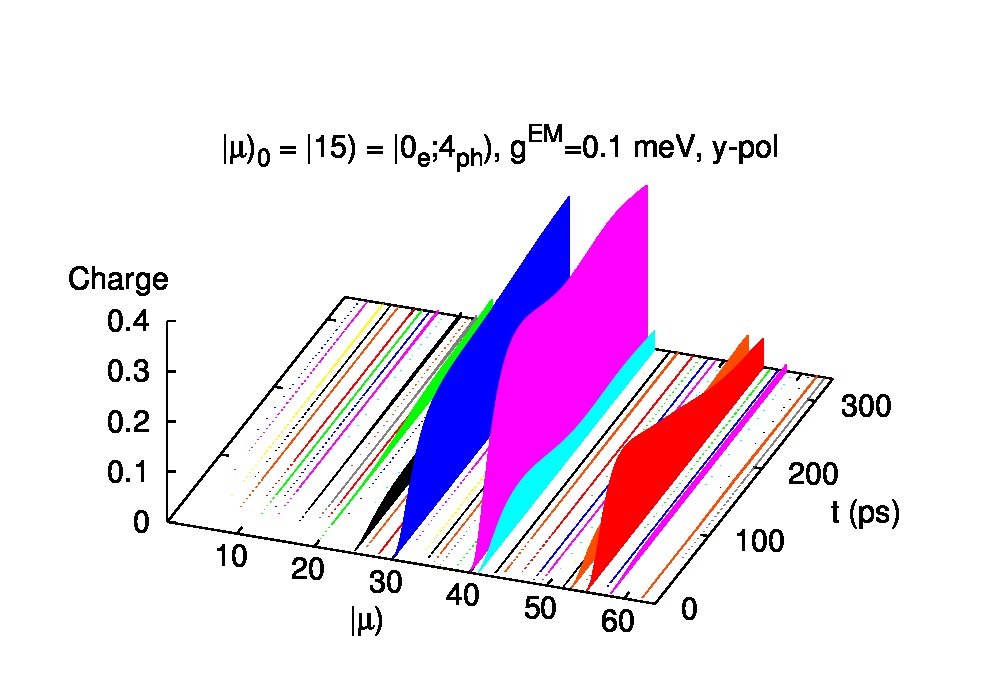}
      \caption{(Color online) The mean number of electrons $\langle N_\mathrm{e}(t)\rangle$ 
               in a MBS $|\breve{\mu})$ for $x$-polarization (left) and $y$-polarization (right)
               as a function of time. Initially, at $t=0$, there is no photon 
               in the cavity (top panels), 1 photon (middle panels), or 4 photons (bottom panels).
               The initial number of electrons is zero in all cases.
               $B=0.1$ T, $g^\mathrm{EM}=0.10$ meV, $\hbar\omega = 0.4$ meV,
               $\mu_L=2.0$ meV, $\mu_R=1.4$ meV,
               $\hbar\Omega_0=1.0$ meV, $\Delta^l_E=0.25$ meV, $g_0^la_w^{3/2}=53.14$ meV,
               $\delta_{1,2}^la_w^2=0.4916$, $L_x=300$ nm, $m^*=0.067m_e$, and $\kappa =12.4$.}
      \label{Net-g010}
\end{figure}

A clearer picture of the charging phenomena can be obtained by observing not the total
amount of charge, but the time-dependent charging of the individual MBSs that are included
in our calculation. This information is presented in Fig.\ \ref{Net-g001} for the weaker
electron-photon coupling, and in Fig.\ \ref{Net-g010} for the stronger one.  

We immediately notice that in the case of the weaker coupling, Fig.\ \ref{Net-g001}, the occupation
of the MBSs is almost independent of the polarization of the cavity photons, but as could be expected 
a higher number of them initially present promotes the occupation of higher energy states. 
The occupation of the MBSs for 4 initial photons is indicative of a system strongly out of
equilibrium with some lower lying MBSs almost empty. 

In the case of the stronger electron-photon coupling the occupation of the MBSs is still
similar to the results for the weaker coupling for the $y$-polarization, but a drastic
change is visible for the $x$-polarization with more MBSs partially occupied.  

\begin{figure}[htbq]
      \includegraphics[width=0.42\textwidth,angle=0,viewport=1 1 346 252,clip]{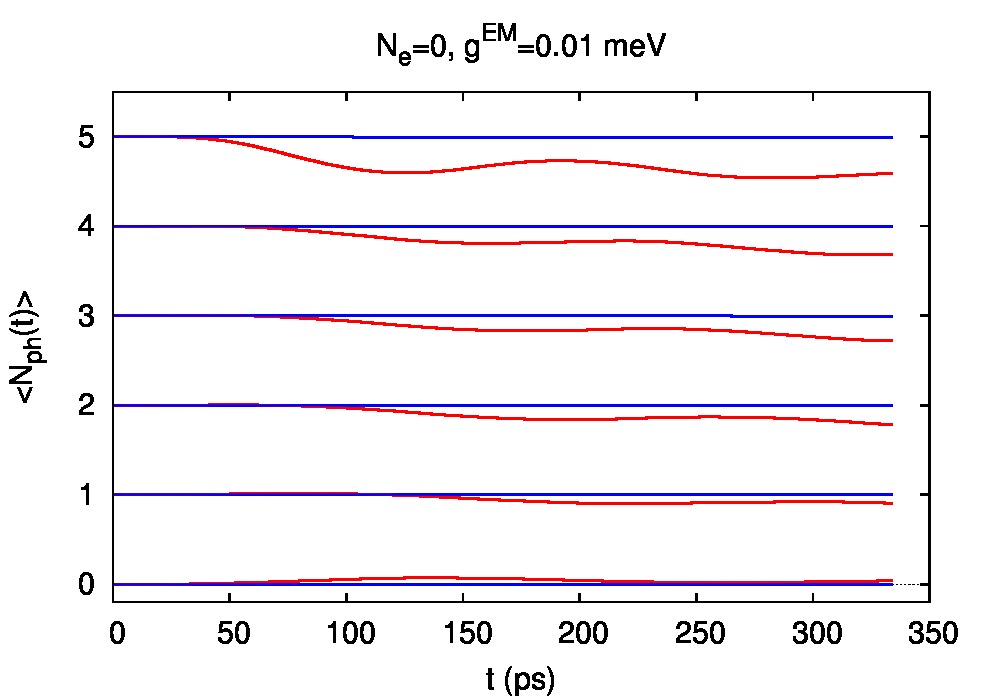}
      \includegraphics[width=0.42\textwidth,angle=0,viewport=1 1 346 252,clip]{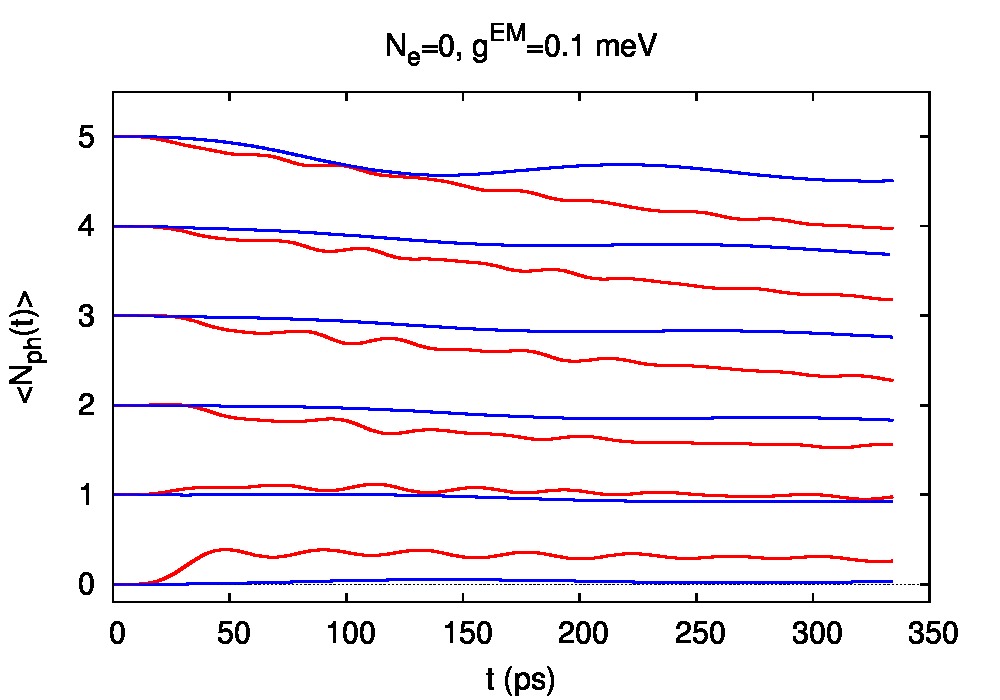}
      \caption{(Color online) The total mean number of photons $\langle N_\mathrm{ph}\rangle$ 
               as a function of time for $g^\mathrm{EM}=0.01$ meV (upper panel) and $g^\mathrm{EM}=0.10$ meV
               (lower panel). $x$-polarization (red), and $y$-polarization (blue).
               $B=0.1$ T, $\hbar\Omega_0=1.0$ meV, $\hbar\omega = 0.4$ meV,
               $\mu_L=2.0$ meV, $\mu_R=1.4$ meV,
               $L_x=300$ nm, $\Delta^l_E=0.25$ meV, $g_0^la_w^{3/2}=53.14$ meV,
               $\delta_{1,2}^la_w^2=0.4916$, $m^*=0.067m_e$, and $\kappa =12.4$.}
      \label{NphtT}
\end{figure}

In order to analyze the effects of different number of photons on the charging of the central
system we show first the time-evolution of the total mean number of photons 
$\langle N_\mathrm{ph}\rangle$ for the two different values of the electron-photon
coupling strength in Fig.\ \ref{NphtT}. In both cases the mean number of photons does
not vary much for the $y$-polarization. Larger deviations are seen for the $x$-polarization
with shorter period oscillations for the stronger coupling. 

\begin{figure}[htbq]
      \includegraphics[width=0.234\textwidth,angle=0,viewport=1 1 325 210,clip]{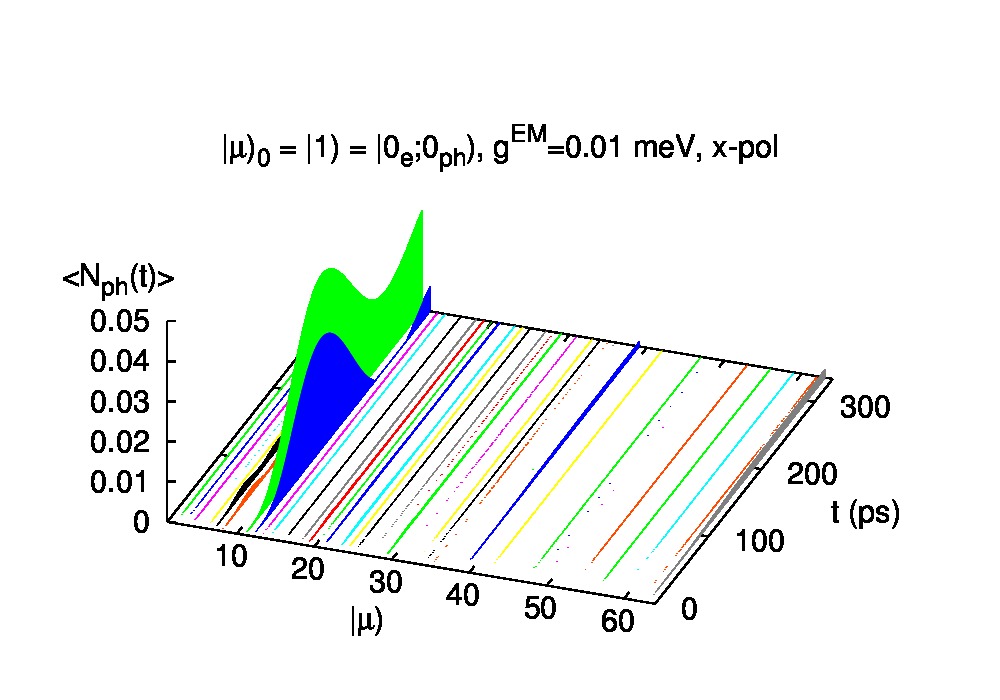}
      \includegraphics[width=0.234\textwidth,angle=0,viewport=1 1 325 210,clip]{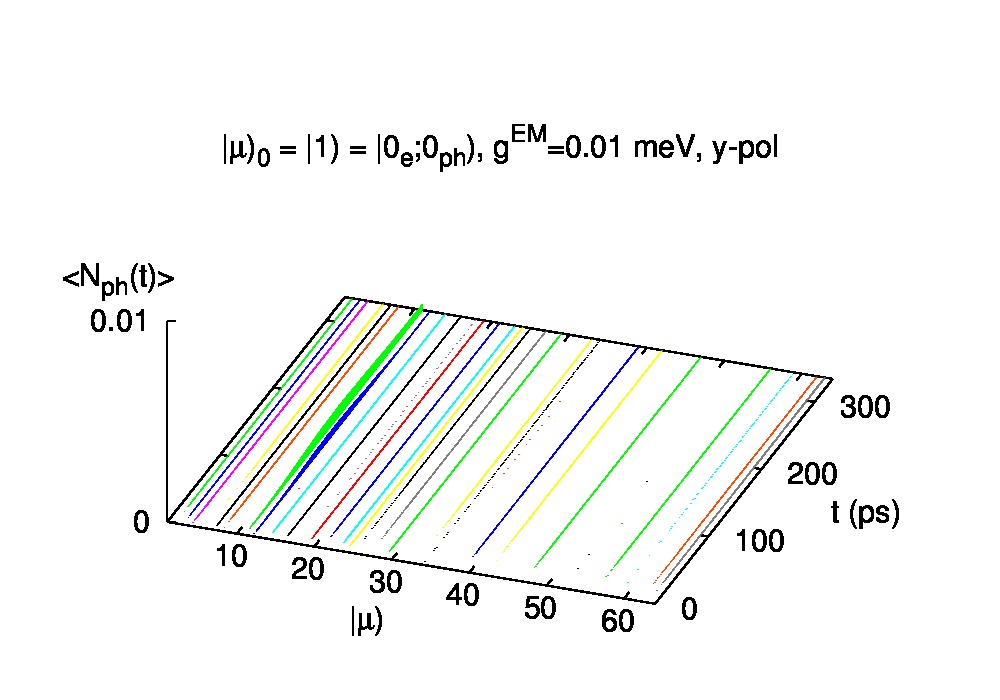}
      \includegraphics[width=0.234\textwidth,angle=0,viewport=1 1 325 210,clip]{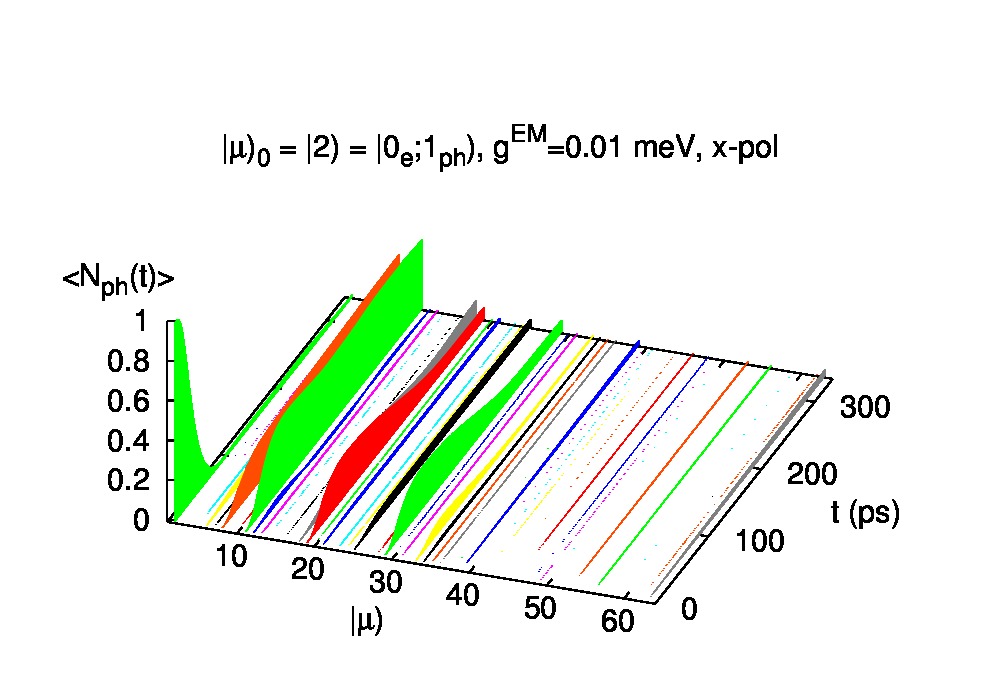}
      \includegraphics[width=0.234\textwidth,angle=0,viewport=1 1 325 210,clip]{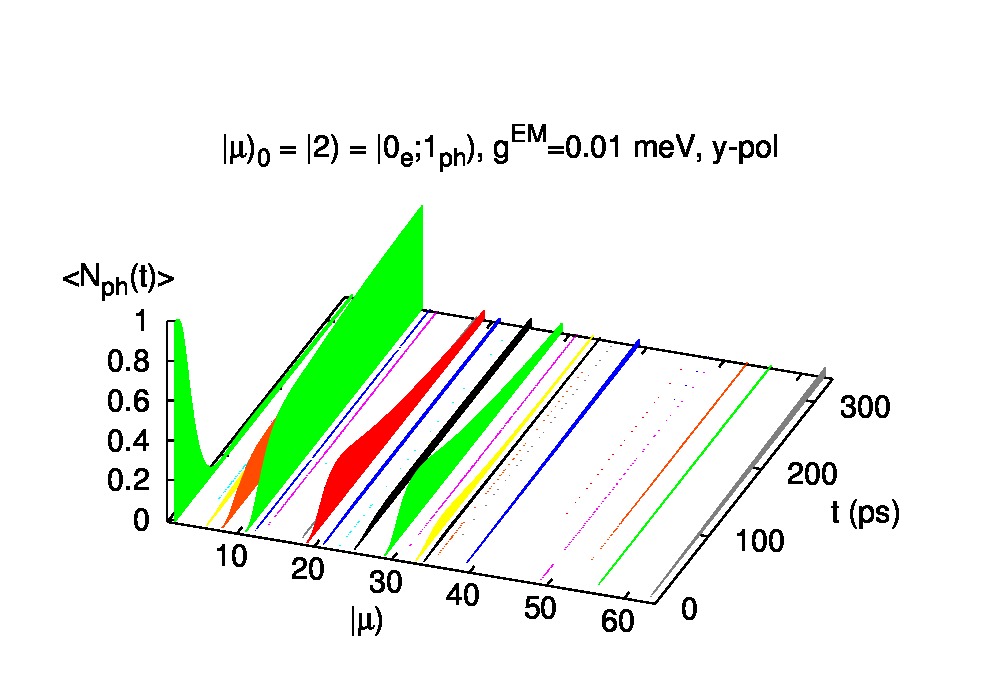}
      \includegraphics[width=0.234\textwidth,angle=0,viewport=1 1 325 210,clip]{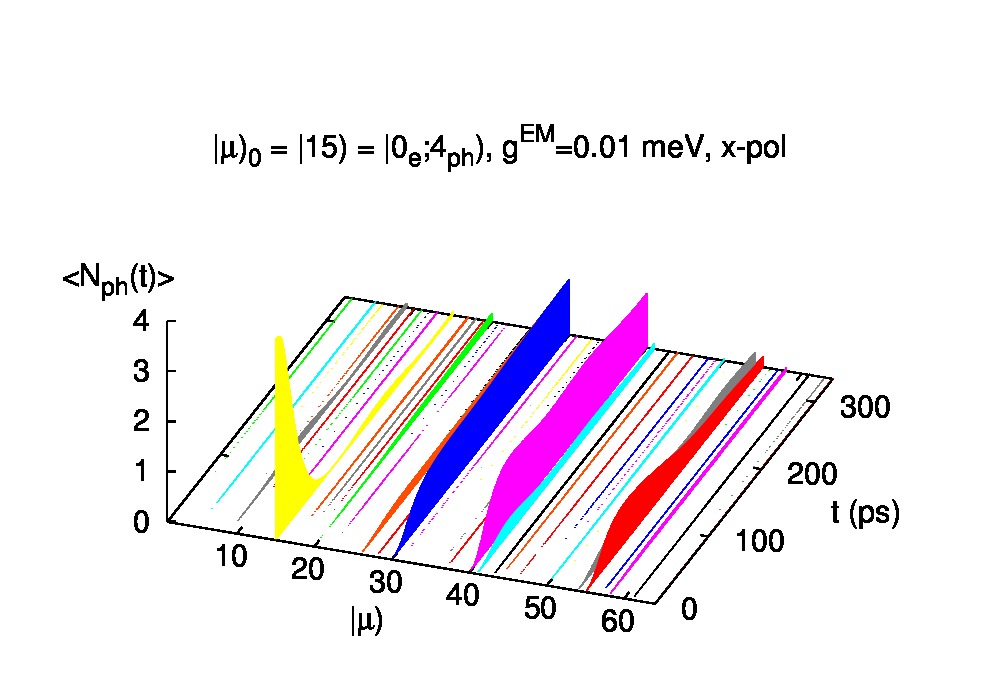}
      \includegraphics[width=0.234\textwidth,angle=0,viewport=1 1 325 210,clip]{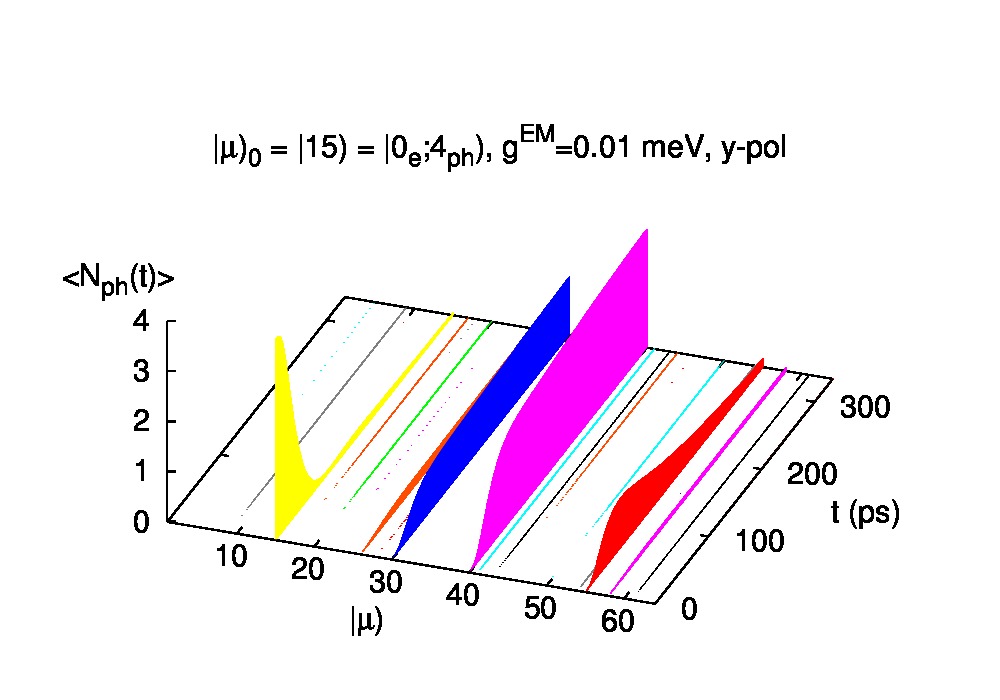}
      \caption{(Color online) The mean number of photons $\langle N_\mathrm{ph}(t)\rangle$ 
               in a MBS $|\breve{\mu})$ for $x$-polarization (left) and $y$-polarization (right)
               as a function of time. Initially, at $t=0$, there are no photons 
               in the cavity (top panels), 1 photon (middle panels), or 4 photons (bottom panels).
               $B=0.1$ T, $g^\mathrm{EM}=0.01$ meV, $\hbar\omega = 0.4$ meV,
               $\mu_L=2.0$ meV, $\mu_R=1.4$ meV,
               $\hbar\Omega_0=1.0$ meV, $\Delta^l_E=0.25$ meV, $g_0^la_w^{3/2}=53.14$ meV,
               $\delta_{1,2}^la_w^2=0.4916$, $L_x=300$ nm, $m^*=0.067m_e$, and $\kappa =12.4$.}
      \label{Npht-g001}
\end{figure}

\begin{figure}[htbq]
      \includegraphics[width=0.234\textwidth,angle=0,viewport=1 1 325 210,clip]{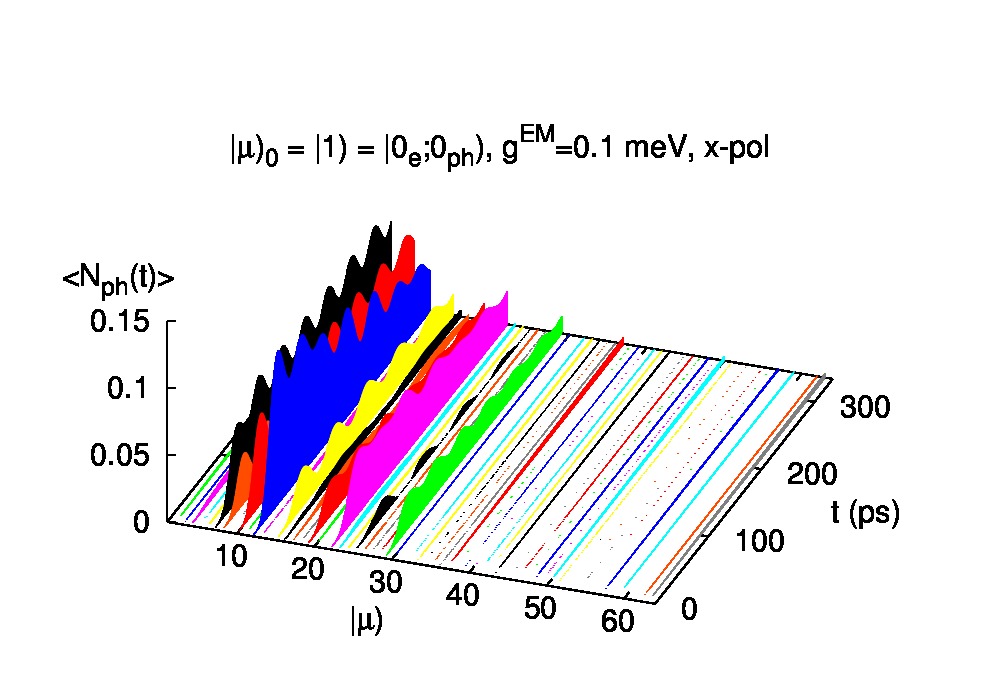}
      \includegraphics[width=0.234\textwidth,angle=0,viewport=1 1 325 210,clip]{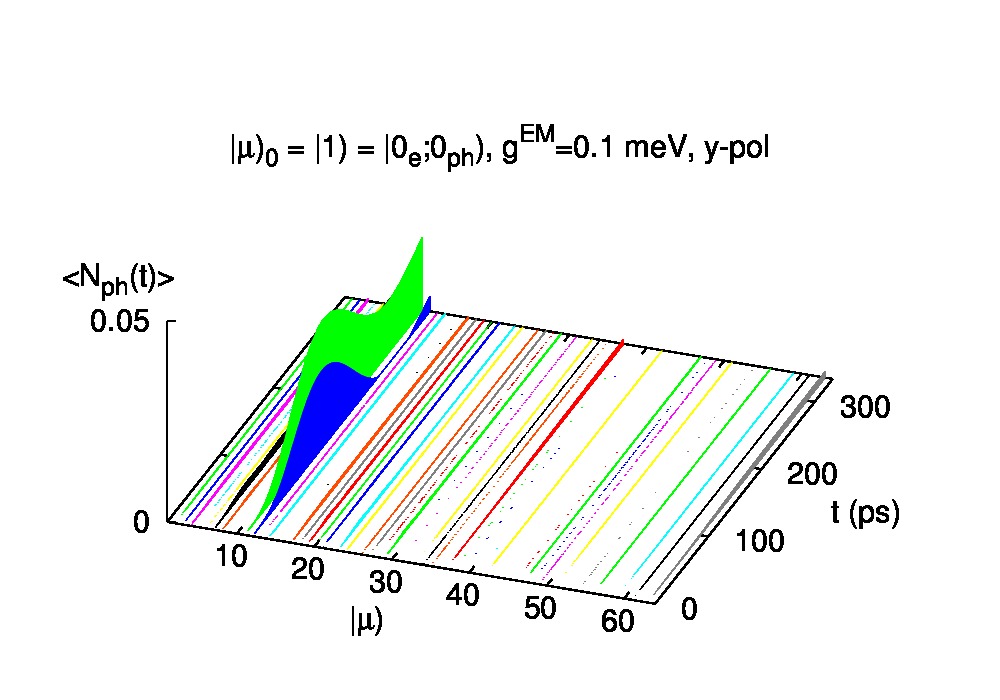}
      \includegraphics[width=0.234\textwidth,angle=0,viewport=1 1 325 210,clip]{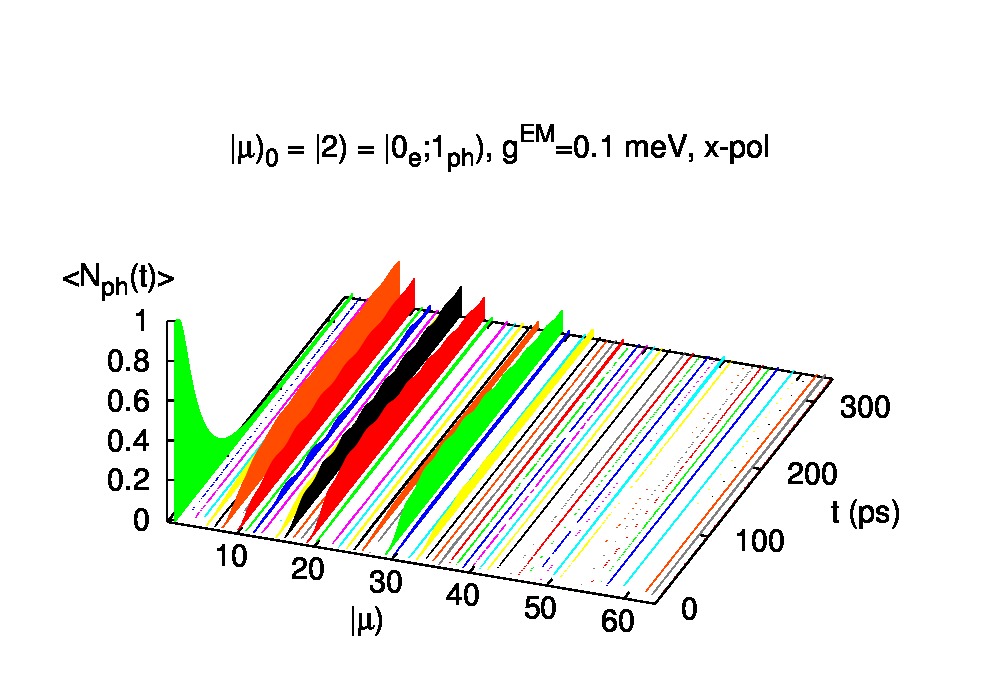}
      \includegraphics[width=0.234\textwidth,angle=0,viewport=1 1 325 210,clip]{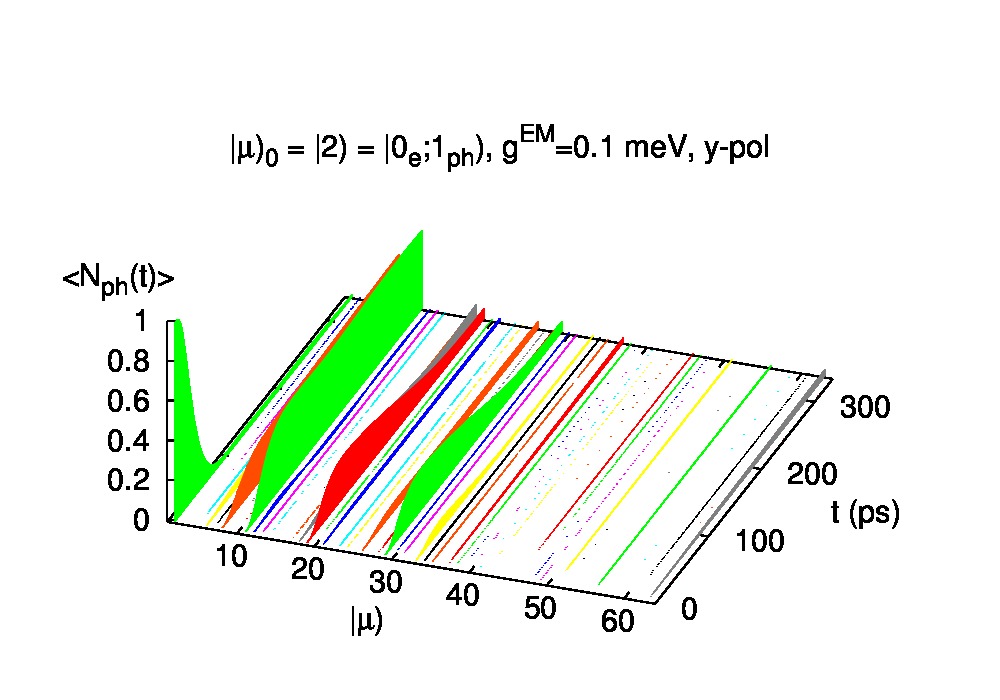}
      \includegraphics[width=0.234\textwidth,angle=0,viewport=1 1 325 210,clip]{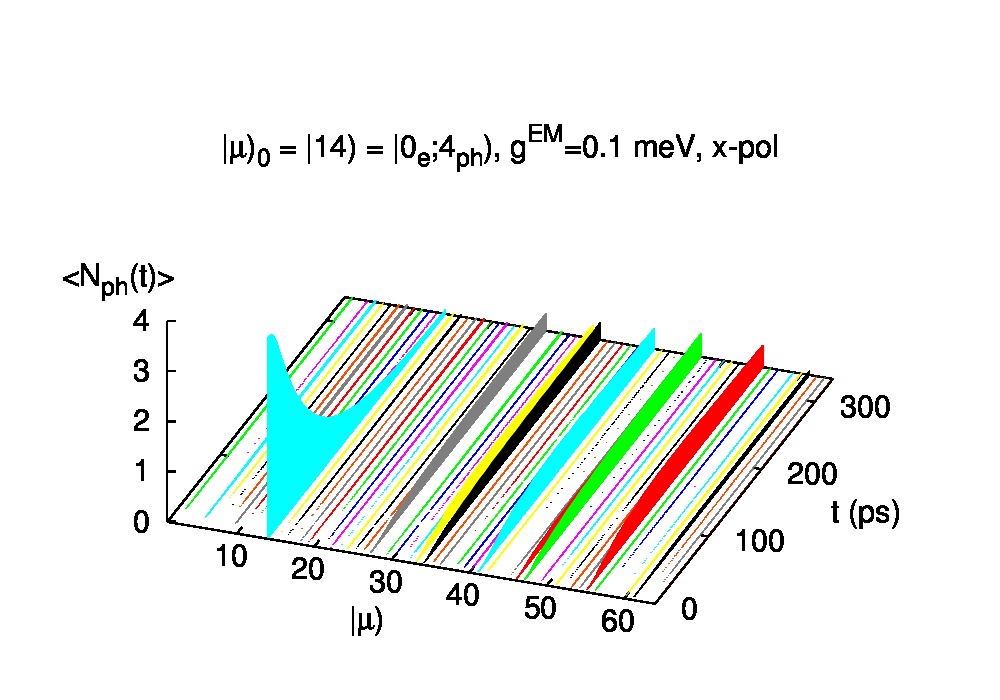}
      \includegraphics[width=0.234\textwidth,angle=0,viewport=1 1 325 210,clip]{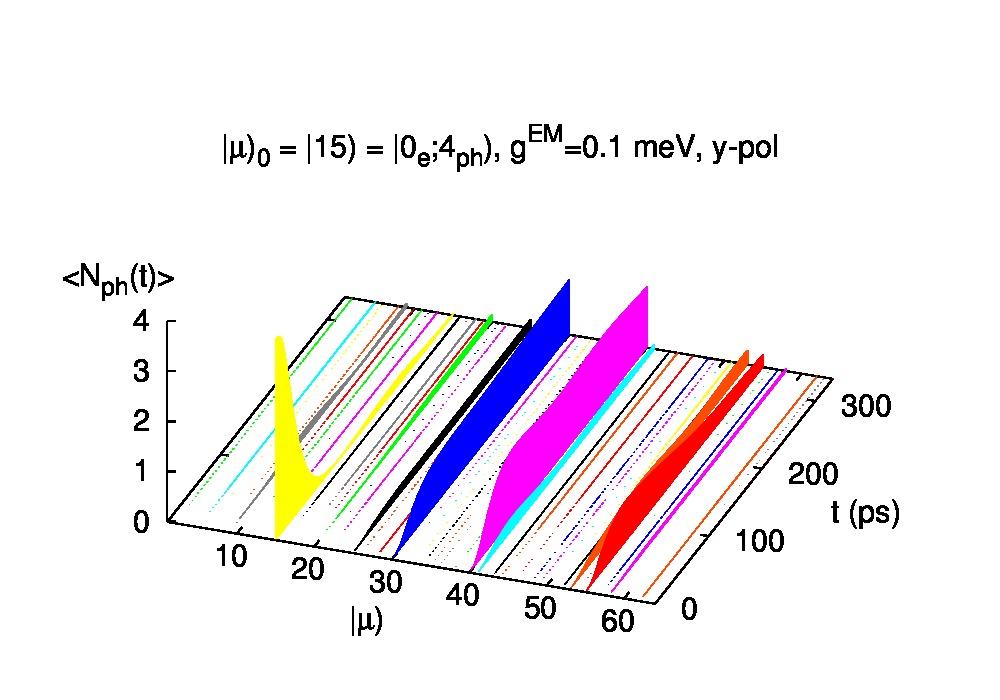}
      \caption{(Color online) The mean number of photons $\langle N_\mathrm{ph}(t)\rangle$ 
               in a MBS $|\breve{\mu})$ for $x$-polarization (left) and $y$-polarization (right)
               as a function of time. Initially, at $t=0$, there are no photons 
               in the cavity (top panels), 1 photon (middle panels), or 4 photons (bottom panels).
               $B=0.1$ T, $g^\mathrm{EM}=0.10$ meV, $\hbar\omega = 0.4$ meV,
               $\mu_L=2.0$ meV, $\mu_R=1.4$ meV,
               $\hbar\Omega_0=1.0$ meV, $\Delta^l_E=0.25$ meV, $g_0^la_w^{3/2}=53.14$ meV,
               $\delta_{1,2}^la_w^2=0.4916$, $L_x=300$ nm, $m^*=0.067m_e$, and $\kappa =12.4$.}
      \label{Npht-g010}
\end{figure}

Again, a better insight into the photon-electron dynamics can be gained by observing the
evolution of the mean photon number of each MBS which is presented in Fig.\ \ref{Npht-g001} for
the weaker coupling and in Fig.\ \ref{Npht-g010} for the stronger one. For both cases
we notice the rapid decay of the photon number from their initial state
if the number was initially 1 or larger. 
With reference and comparison to Fig.\ \ref{N-virkjar} we observe a fast build up of
electron-photon many-body states with a considerable photon component and higher
energy. The total number of photons does not change fast in the system, but the introduction
of electrons to the system through the leads results in a fast redistribution of the
photons into many-body states of quasiparticles, or in other words the preexisting 
initial photons participate in and facilitate the build up of many-body states with higher energy.  
The slow change in the total photon number in Fig.\ \ref{NphtT} reflects phenomena of 
radiation absorption and emission while the rapid change with time in the occupation
in Figs \ref{Net-g001} and \ref{Net-g010} indicates the creation of quasiparticles
with definite electron and photon components. We thus observe phenomena on at least
two different time scales.  

We notice again (see Fig.\ \ref{Npht-g001}) that for the weak electron-photon coupling 
there is not a large difference
between the two polarizations except in the case of no initial photon. 

It is well known that the GME in the approximation used here can lead to nonphysical 
negative probability for occupation of individual many-body levels if the coupling to
the reservoirs is too strong.\cite{Whitney08:175304} 
We have avoided this problem by keeping the coupling low
enough. It is interesting to note here that it is our experience that after coupling to 
the photon system such a negative probability will first turn up in the probability for 
a photon occupation before it would be seen in the probability for the occpation of an 
electron.  

\begin{figure}[htbq]
      \includegraphics[width=0.40\textwidth,angle=0,viewport=1 1 346 252,clip]{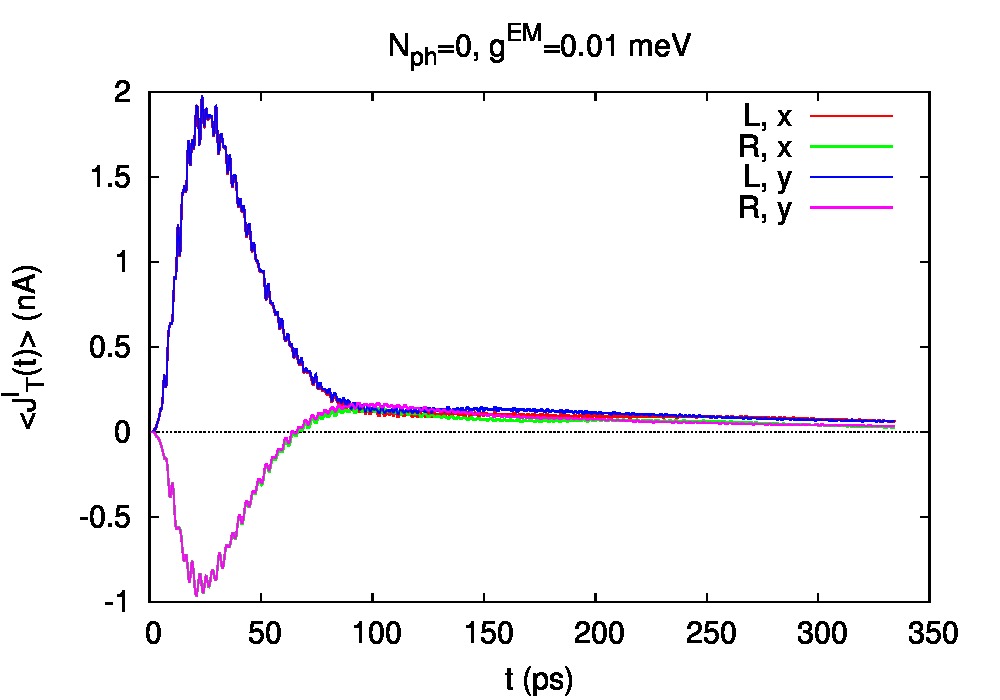}
      \includegraphics[width=0.40\textwidth,angle=0,viewport=1 1 346 252,clip]{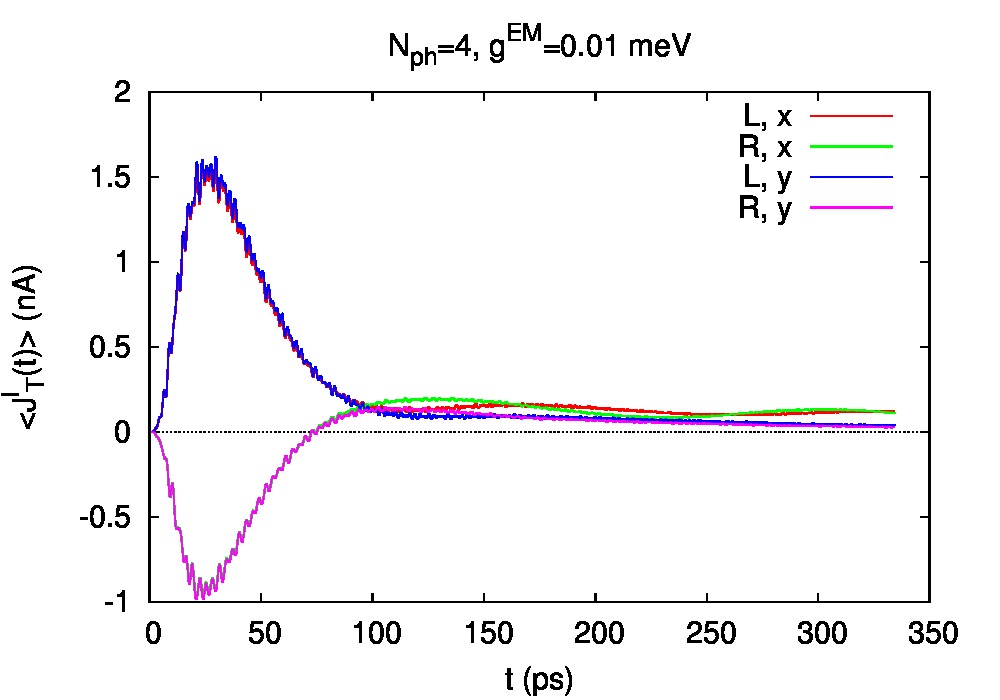}
      \caption{(Color online) The total current from the left lead (L), and into the right lead (R) 
               as a function of time for $g^\mathrm{EM}=0.01$ meV. 
               Initially, at $t=0$ there are no photons in the cavity (top panel), or
               4 photons (bottom panel). 
               $B=0.1$ T, $\hbar\Omega_0=1.0$ meV, $L_x=300$ nm, $\hbar\omega = 0.4$ meV,
               $\mu_L=2.0$ meV, $\mu_R=1.4$ meV, $\Delta^l_E=0.25$ meV, $g_0^la_w^{3/2}=53.14$ meV,
               $\delta_{1,2}^la_w^2=0.4916$,
               $m^*=0.067m_e$, and $\kappa =12.4$.}
      \label{Jtxy-g001}
\end{figure}

\begin{figure}[htbq]
      \includegraphics[width=0.40\textwidth,angle=0,viewport=1 1 346 252,clip]{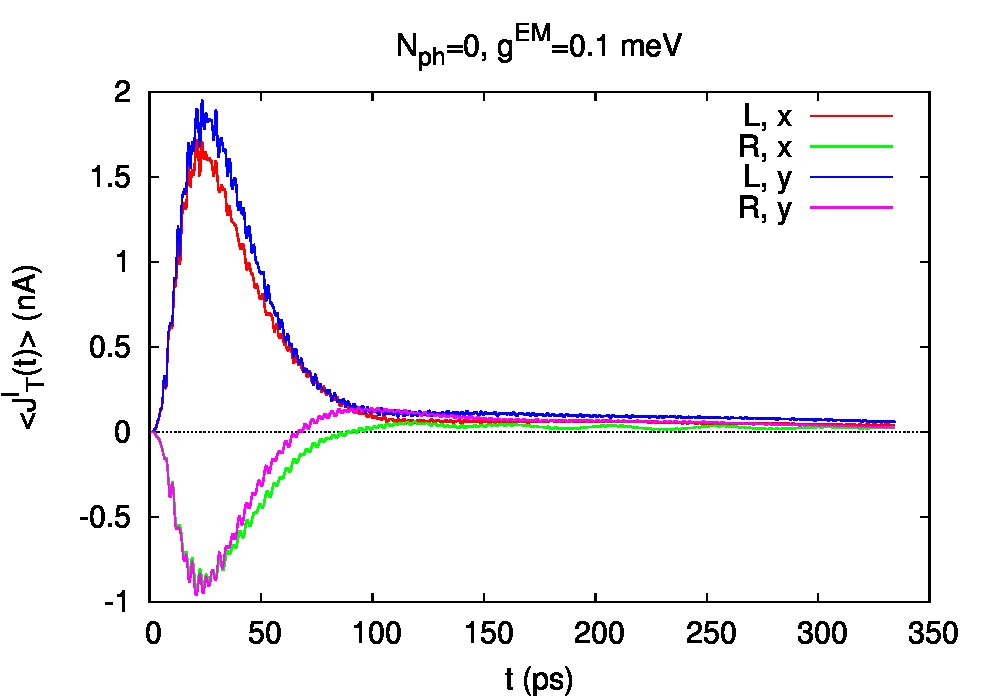}
      \includegraphics[width=0.40\textwidth,angle=0,viewport=1 1 346 252,clip]{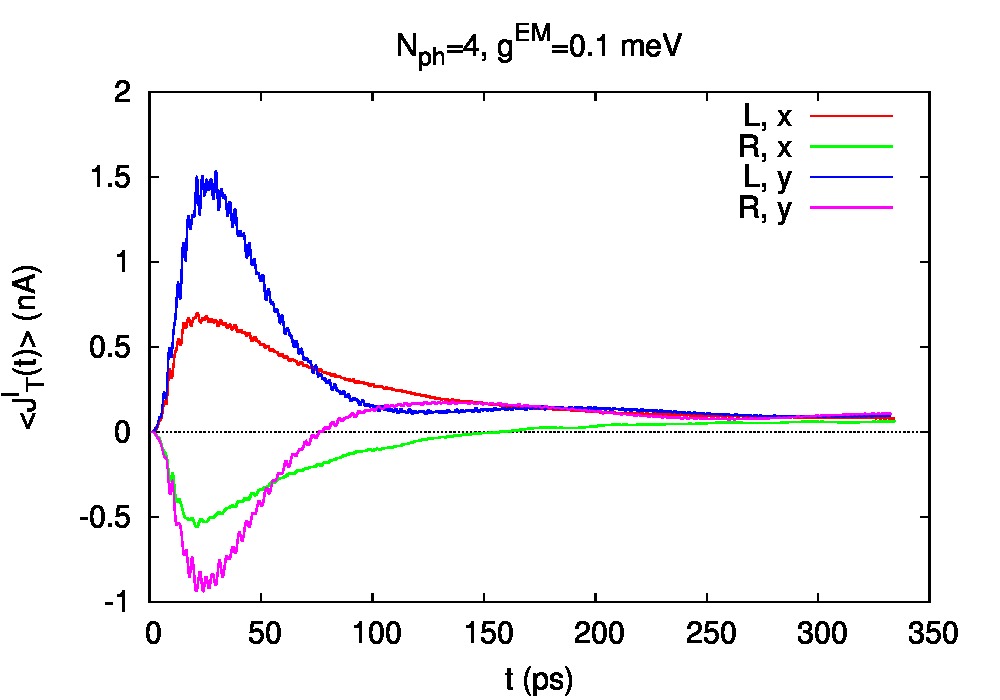}
      \caption{(Color online) The total current from the left lead (L), and into the right lead (R) 
               as a function of time for $g^\mathrm{EM}=0.1$ meV. Initially, at $t=0$ there are no photons 
               in the cavity (top panel), or
               4 photons (bottom panel). 
               $B=0.1$ T, $\hbar\Omega_0=1.0$ meV, $L_x=300$ nm, $\hbar\omega = 0.4$ meV,
               $\mu_L=2.0$ meV, $\mu_R=1.4$ meV, $\Delta^l_E=0.25$ meV, $g_0^la_w^{3/2}=53.14$ meV,
               $\delta_{1,2}^la_w^2=0.4916$,
               $m^*=0.067m_e$, and $\kappa =12.4$.}
      \label{Jtxy-g010}
\end{figure}

The total current from the left lead into the system and from the system to the
right lead is displayed in Figs \ref{Jtxy-g001} and \ref{Jtxy-g010} for both polarizations.
The irregular short period oscillation seen in the current stems from addition of 
partial currents per each many-body state with simple almost harmonic oscillations of various
short periods characteristic for each component. 
For the weak coupling, $g^\mathrm{EM}=0.01$ meV, and no photon present at $t=0$ 
the left- and right currents are almost identical and very close to the current
for the electron system with no coupling to cavity photons. This changes slightly
for some photons initially present in the system, especially for $t\sim 300$ ps when the
system is approaching a steady state. Here, we should mention that the
photon energy, $\hbar\omega = 0.4$ meV, was selected to be smaller than the 
characteristic confinement energy in the $y$-direction and not to be in resonance
with any Coulomb interacting MESs.

In the case of a stronger electron-photon coupling, $g^\mathrm{EM}=0.1$ meV, 
the charging of the central system is attenuated by the presence of $x$-polarized photons
at $t=0$. We are describing a central system here in the absence of any potential 
that can give its eigenstates different localization character, see Fig.\ \ref{JE},
we do thus not expect any simple phenomena of photo enhanced conduction.
The photon energy $\hbar\omega = 0.4$ meV is far from being in resonance with the
characteristic energy for the $y$-confinement, but it is much closer to the 
characteristic energy for the $x$-direction. Inspection of the charge distribution
of, or the occupation of many-body states $|\breve{\mu})$ in Fig.\ \ref{Net-g010}
and the distribution of the photons in the same states in Fig.\ \ref{Npht-g010}
demonstrates a strong correlation between the states and that the initial photons
have caused the electrons to be distributed into many states. In a linear response
situation which we do not have here we would say that the photon scattering of the
electrons reduced the conduction. Here we have to say that the stronger photon-electron 
interaction for the case of the $x$-polarization attenuates the charging of the
central system. Observation of the partial currents through the many-body state, 
not shown here, tells the same story. There are more states contributing to the
charging, but they all bring less charge into the system. To understand this
situation it is good to have in mind the similarity between the cavity photons
here and confined phonons.

\section{Summary}
In this article we have taken the first steps to describe the transport of Coulomb interacting 
electrons through a photon cavity taking into account the geometry of the central system
and the leads and allowing for a strong coupling to the photon mode.
The central system is a parabolic quantum wire of finite length. Its aspect ratio and
confinement characteristics make excitations for low values of the external magnetic field
easier in the transport direction, the $x$-direction, than perpendicular to it, in the $y$-direction.                            
By selecting the energy of the photon mode below the characteristic confinement frequency in the
$y$-direction we demonstrate that the transport becomes very dependent on the polarization of the
cavity mode and the number of photons initially present in the system at $t=0$, before the 
electrons enter it. Generally, a higher initial photon number reduces the transient charging 
of the system with electrons. This effect is enhanced by increased coupling between the electrons
and the photons. The largest reduction is found for several $x$-polarized photons in the system.
The photon energy is then high enough to disperse the electrons into numerous excited states.
The system does only contain delocalized electron states and this behavior has resemblance to
reduced conductance due to phonon scattering of electrons. It has though to be kept in mind that
we are observing a transient behavior here in the charging phase of the central system. 

We observe dynamics on two different time scales here, the fast dispersion of electrons and
photons into excited states initially seen in the transient regime, especially when analyzed
for each many-body state, see Figs \ref{Net-g001}, \ref{Net-g010}, 
\ref{Npht-g001}, and \ref{Npht-g010}, and the slow decay or gain caused by radiation phenomena and 
reflected by the total number of photons in Fig.\ \ref{NphtT}. 
Connected to the issue of the two time scales it is interesting to notice that in the
case of no photon present in the system initially only many-body states with low energy and almost
vanishing photon content are occupied by the electron entering the system. In this situation
the total photon content of the system only changes on the radiation time scale.

In the leads we have only electrons and in the central system quasiparticles with photon
and electron content. Their lifetime is determined by the coupling to the leads and the time scale
of radiation processes. The original coupling Hamiltonian of the leads and the central
system, $H_\mathrm{T}$, describes the entry or exit of noninteracting electrons to the
central system from the leads. The two unitary transformations of $H_\mathrm{T}$, first
to the basis of interacting MESs and the second one to the electron-photon many-body basis
guarantee that the coupling between the leads and the central system describes how 
electrons leave or enter the leads, and conversely how the number of quasiparticles or photons
changes in the system. This stepwise introduction of complexity to the model, concurrent stepwise
truncation of the ensuing Fock-space, and unitary transformation of the Hamiltonian at each
step to the appropriate basis enables us to attack a problem that otherwise requires too
large Fock space for numerical calculations. This proceedure can still be streamlined in
order to describe systems with stronger electron-photon coupling or more complex geometry.  

The anisotropic response of the electron system to the two different polarizations of the
photon field is caused by the confinement energy in the $y$-direction, 
$\hbar\Omega_0= 1.0$ meV, beeing larger than both the photon energy, $\hbar\omega = 0.4$ meV,
and the lowest excitation in the $x$-direction, $\Delta E_x\sim 0.2$ meV. At the low energy
we are observing the properties of the system here it is much ``harder'' or ``stiffer'' in the 
$y$- than the $x$-direction. We thus do not see much change in the properties of the system
(see Fig.\ \ref{Jtxy-g001} and \ref{Jtxy-g010}) varying the initial photon number or 
the coupling strength to the photons for the parameter range explored here. 
For the low coupling strength, $g^\mathrm{EM}=0.01$ meV
there is not much difference either between the system properties for the two polarizations,
but it is interesting to notice how similar the charging per MBS for the stronger coupling,
$g^\mathrm{EM}=0.1$ meV, and $y$-polarization (see the right panels of Fig.\ \ref{Net-g010})
is to the charging at the weaker coupling and $x$-polarization 
(see the left panels of Fig.\ \ref{Net-g001}). The same analogy is found for the 
photon component in the MBS (compare the right panels of Fig.\ \ref{Npht-g010} to the
left panels of Fig.\ \ref{Npht-g001}). This analogy hints at a highly anisotropic
``effective coupling'' of the electrons and cavity photons for the geometry selectrd here.
 
The strong coupling of the electrons to the photon field leads to polarization effects in our
system that require a large functional basis for the electron states of our system to describe
correctly. This together with other important properties of the closed system will be discussed
elsewhere\cite{Jonasson2011:01} and is the limiting factor here in how strong coupling we can 
describe in our present system and still manage observation of the time evolution according to the GME.

%
%
\begin{acknowledgments}
      The authors acknowledge discussions with Ivan Shelykh
      and Valeriu Moldoveanu.
      The authors acknowledge financial support from the Icelandic Research
      and Instruments Funds,
      the Research Fund of the University of Iceland, the
      National Science Council of Taiwan under contract
      No.\ NSC100-2112-M-239-001-MY3. HSG acknowledges support from the National Science Council, 
      Taiwan, under Grants No.\ 97-2112-M-002-012-MY3, and No.\ 100-2112-M-002-003-MY3, 
      support from the Frontier and Innovative Research Program of the National Taiwan University 
      under Grants No.\ 99R80869 and No. 99R80871, and support from the
      focus group program of the National Center for Theoretical Sciences, Taiwan.

\end{acknowledgments}
%
%
%
\bibliographystyle{apsrev4-1}

\begin{thebibliography}{42}%
\makeatletter
\providecommand \@ifxundefined [1]{%
 \@ifx{#1\undefined}
}%
\providecommand \@ifnum [1]{%
 \ifnum #1\expandafter \@firstoftwo
 \else \expandafter \@secondoftwo
 \fi
}%
\providecommand \@ifx [1]{%
 \ifx #1\expandafter \@firstoftwo
 \else \expandafter \@secondoftwo
 \fi
}%
\providecommand \natexlab [1]{#1}%
\providecommand \enquote  [1]{``#1''}%
\providecommand \bibnamefont  [1]{#1}%
\providecommand \bibfnamefont [1]{#1}%
\providecommand \citenamefont [1]{#1}%
\providecommand \href@noop [0]{\@secondoftwo}%
\providecommand \href [0]{\begingroup \@sanitize@url \@href}%
\providecommand \@href[1]{\@@startlink{#1}\@@href}%
\providecommand \@@href[1]{\endgroup#1\@@endlink}%
\providecommand \@sanitize@url [0]{\catcode `\\12\catcode `\$12\catcode
  `\&12\catcode `\#12\catcode `\^12\catcode `\_12\catcode `\%12\relax}%
\providecommand \@@startlink[1]{}%
\providecommand \@@endlink[0]{}%
\providecommand \url  [0]{\begingroup\@sanitize@url \@url }%
\providecommand \@url [1]{\endgroup\@href {#1}{\urlprefix }}%
\providecommand \urlprefix  [0]{URL }%
\providecommand \Eprint [0]{\href }%
\@ifxundefined \urlstyle {%
  \providecommand \doi  [0]{\begingroup \@sanitize@url \@doi}%
  \providecommand \@doi [1]{\endgroup \@@startlink {\doibase
  #1}doi:\discretionary {}{}{}#1\@@endlink }%
}{%
  \providecommand \doi  [0]{doi:\discretionary{}{}{}\begingroup
  \urlstyle{rm}\Url }%
}%
\providecommand \doibase [0]{http://dx.doi.org/}%
\providecommand \Doi [0]{\begingroup \@sanitize@url \@Doi }%
\providecommand \@Doi  [1]{\endgroup\@@startlink{\doibase#1}\@@Doi}%
\providecommand \@@Doi [1]{#1\@@endlink}%
\providecommand \selectlanguage [0]{\@gobble}%
\providecommand \bibinfo  [0]{\@secondoftwo}%
\providecommand \bibfield  [0]{\@secondoftwo}%
\providecommand \translation [1]{[#1]}%
\providecommand \BibitemOpen [0]{}%
\providecommand \bibitemStop [0]{}%
\providecommand \bibitemNoStop [0]{.\EOS\space}%
\providecommand \EOS [0]{\spacefactor3000\relax}%
\providecommand \BibitemShut  [1]{\csname bibitem#1\endcsname}%
\bibitem [{\citenamefont {Tang}\ and\ \citenamefont {Chu}(1999)}]{Tang99:1830}%
  \BibitemOpen
  \bibfield  {author} {\bibinfo {author} {\bibfnamefont {C.~S.}\ \bibnamefont
  {Tang}}\ and\ \bibinfo {author} {\bibfnamefont {C.~S.}\ \bibnamefont {Chu}},\
  }\Doi {10.1103/PhysRevB.60.1830} {\bibfield  {journal} {\bibinfo  {journal}
  {Phys. Rev. B},\ }\textbf {\bibinfo {volume} {60}},\ \bibinfo {pages} {1830}
  (\bibinfo {year} {1999})}\BibitemShut {NoStop}%
\bibitem [{\citenamefont {Tang}\ \emph {et~al.}(2003)\citenamefont {Tang},
  \citenamefont {Tan},\ and\ \citenamefont {Chu}}]{Tang03:205324}%
  \BibitemOpen
  \bibfield  {author} {\bibinfo {author} {\bibfnamefont {C.~S.}\ \bibnamefont
  {Tang}}, \bibinfo {author} {\bibfnamefont {Y.~H.}\ \bibnamefont {Tan}}, \
  and\ \bibinfo {author} {\bibfnamefont {C.~S.}\ \bibnamefont {Chu}},\
  }\href@noop {} {\bibfield  {journal} {\bibinfo  {journal} {Phys. Rev. B},\
  }\textbf {\bibinfo {volume} {67}},\ \bibinfo {pages} {205324} (\bibinfo
  {year} {2003})}\BibitemShut {NoStop}%
\bibitem [{\citenamefont {Zhou}\ and\ \citenamefont {Li}(2005)}]{Zhou05:6663}%
  \BibitemOpen
  \bibfield  {author} {\bibinfo {author} {\bibfnamefont {G.}~\bibnamefont
  {Zhou}}\ and\ \bibinfo {author} {\bibfnamefont {Y.}~\bibnamefont {Li}},\
  }\Doi {10.1088/0953-8984/17/42/007} {\bibfield  {journal} {\bibinfo
  {journal} {J. Phys.: Condens. Matter},\ }\textbf {\bibinfo {volume} {17}},\
  \bibinfo {pages} {6663} (\bibinfo {year} {2005})}\BibitemShut {NoStop}%
\bibitem [{\citenamefont {Zeb}\ \emph {et~al.}(2008)\citenamefont {Zeb},
  \citenamefont {Sabeeh},\ and\ \citenamefont {Tahir}}]{Zeb08:165420}%
  \BibitemOpen
  \bibfield  {author} {\bibinfo {author} {\bibfnamefont {M.~A.}\ \bibnamefont
  {Zeb}}, \bibinfo {author} {\bibfnamefont {K.}~\bibnamefont {Sabeeh}}, \ and\
  \bibinfo {author} {\bibfnamefont {M.}~\bibnamefont {Tahir}},\ }\Doi
  {10.1103/PhysRevB.78.165420} {\bibfield  {journal} {\bibinfo  {journal}
  {Phys. Rev B},\ }\textbf {\bibinfo {volume} {78}},\ \bibinfo {pages} {165420}
  (\bibinfo {year} {2008})}\BibitemShut {NoStop}%
\bibitem [{\citenamefont {Lin}\ \emph {et~al.}(2008)\citenamefont {Lin},
  \citenamefont {Tang},\ and\ \citenamefont {Chang}}]{Lin08:245312}%
  \BibitemOpen
  \bibfield  {author} {\bibinfo {author} {\bibfnamefont {C.~H.}\ \bibnamefont
  {Lin}}, \bibinfo {author} {\bibfnamefont {C.~S.}\ \bibnamefont {Tang}}, \
  and\ \bibinfo {author} {\bibfnamefont {Y.~C.}\ \bibnamefont {Chang}},\ }\Doi
  {10.1103/PhysRevB.78.245312} {\bibfield  {journal} {\bibinfo  {journal}
  {Phys. Rev. B},\ }\textbf {\bibinfo {volume} {78}},\ \bibinfo {pages}
  {245312} (\bibinfo {year} {2008})}\BibitemShut {NoStop}%
\bibitem [{\citenamefont {Torfason}\ \emph {et~al.}(2009)\citenamefont
  {Torfason}, \citenamefont {Tang},\ and\ \citenamefont
  {Gudmundsson}}]{Torfason09:195322}%
  \BibitemOpen
  \bibfield  {author} {\bibinfo {author} {\bibfnamefont {K.}~\bibnamefont
  {Torfason}}, \bibinfo {author} {\bibfnamefont {C.-S.}\ \bibnamefont {Tang}},
  \ and\ \bibinfo {author} {\bibfnamefont {V.}~\bibnamefont {Gudmundsson}},\
  }\Doi {10.1103/PhysRevB.80.195322} {\bibfield  {journal} {\bibinfo  {journal}
  {Phys. Rev. B},\ }\textbf {\bibinfo {volume} {80}},\ \bibinfo {pages}
  {195322} (\bibinfo {year} {2009})}\BibitemShut {NoStop}%
\bibitem [{\citenamefont {Kienle}\ and\ \citenamefont
  {L{\'e}onard}(2009)}]{Kienle09:026601}%
  \BibitemOpen
  \bibfield  {author} {\bibinfo {author} {\bibfnamefont {D.}~\bibnamefont
  {Kienle}}\ and\ \bibinfo {author} {\bibfnamefont {F.}~\bibnamefont
  {L{\'e}onard}},\ }\Doi {10.1103/PhysRevLett.103.026601} {\bibfield  {journal}
  {\bibinfo  {journal} {Phys. Rev. Lett.},\ }\textbf {\bibinfo {volume}
  {103}},\ \bibinfo {pages} {026601} (\bibinfo {year} {2009})}\BibitemShut
  {NoStop}%
\bibitem [{\citenamefont {My{\"o}h{\"a}nen}\ \emph {et~al.}(2009)\citenamefont
  {My{\"o}h{\"a}nen}, \citenamefont {Stan}, \citenamefont {Stefanucci},\ and\
  \citenamefont {van Leeuwen}}]{Myohanen09:115107}%
  \BibitemOpen
  \bibfield  {author} {\bibinfo {author} {\bibfnamefont {P.}~\bibnamefont
  {My{\"o}h{\"a}nen}}, \bibinfo {author} {\bibfnamefont {A.}~\bibnamefont
  {Stan}}, \bibinfo {author} {\bibfnamefont {G.}~\bibnamefont {Stefanucci}}, \
  and\ \bibinfo {author} {\bibfnamefont {R.}~\bibnamefont {van Leeuwen}},\
  }\href {http://link.aps.org/doi/10.1103/PhysRevB.80.115107} {\bibfield
  {journal} {\bibinfo  {journal} {Phys. Rev. B},\ }\textbf {\bibinfo {volume}
  {80}},\ \bibinfo {pages} {115107} (\bibinfo {year} {2009})}\BibitemShut
  {NoStop}%
\bibitem [{\citenamefont {Stefanucci}\ \emph {et~al.}(2010)\citenamefont
  {Stefanucci}, \citenamefont {Perfetto},\ and\ \citenamefont
  {Cini}}]{Stefanucci10:115446}%
  \BibitemOpen
  \bibfield  {author} {\bibinfo {author} {\bibfnamefont {G.}~\bibnamefont
  {Stefanucci}}, \bibinfo {author} {\bibfnamefont {E.}~\bibnamefont
  {Perfetto}}, \ and\ \bibinfo {author} {\bibfnamefont {M.}~\bibnamefont
  {Cini}},\ }\Doi {10.1103/PhysRevB.81.115446} {\bibfield  {journal} {\bibinfo
  {journal} {Phys. Rev. B},\ }\textbf {\bibinfo {volume} {81}},\ \bibinfo
  {pages} {115446} (\bibinfo {year} {2010})}\BibitemShut {NoStop}%
\bibitem [{\citenamefont {Abdullah}\ \emph {et~al.}(2010)\citenamefont
  {Abdullah}, \citenamefont {Tang},\ and\ \citenamefont
  {Gudmundsson}}]{Abdullah10:195325}%
  \BibitemOpen
  \bibfield  {author} {\bibinfo {author} {\bibfnamefont {N.~R.}\ \bibnamefont
  {Abdullah}}, \bibinfo {author} {\bibfnamefont {C.-S.}\ \bibnamefont {Tang}},
  \ and\ \bibinfo {author} {\bibfnamefont {V.}~\bibnamefont {Gudmundsson}},\
  }\Doi {10.1103/PhysRevB.82.195325} {\bibfield  {journal} {\bibinfo  {journal}
  {Phys. Rev. B},\ }\textbf {\bibinfo {volume} {82}},\ \bibinfo {pages}
  {195325} (\bibinfo {year} {2010})}\BibitemShut {NoStop}%
\bibitem [{\citenamefont {Tahir}\ and\ \citenamefont
  {MacKinnon}(2010)}]{Tahir10:195444}%
  \BibitemOpen
  \bibfield  {author} {\bibinfo {author} {\bibfnamefont {M.}~\bibnamefont
  {Tahir}}\ and\ \bibinfo {author} {\bibfnamefont {A.}~\bibnamefont
  {MacKinnon}},\ }\Doi {10.1103/PhysRevB.81.195444} {\bibfield  {journal}
  {\bibinfo  {journal} {Phys. Rev. B},\ }\textbf {\bibinfo {volume} {81}},\
  \bibinfo {pages} {195444} (\bibinfo {year} {2010})}\BibitemShut {NoStop}%
\bibitem [{\citenamefont {Chen}\ \emph {et~al.}(2011)\citenamefont {Chen},
  \citenamefont {Jian},\ and\ \citenamefont {Goan}}]{Chen11:115439}%
  \BibitemOpen
  \bibfield  {author} {\bibinfo {author} {\bibfnamefont {P.-W.}\ \bibnamefont
  {Chen}}, \bibinfo {author} {\bibfnamefont {C.-C.}\ \bibnamefont {Jian}}, \
  and\ \bibinfo {author} {\bibfnamefont {H.-S.}\ \bibnamefont {Goan}},\ }\Doi
  {10.1103/PhysRevB.83.115439} {\bibfield  {journal} {\bibinfo  {journal}
  {Phys. Rev. B},\ }\textbf {\bibinfo {volume} {83}},\ \bibinfo {pages}
  {115439} (\bibinfo {year} {2011})}\BibitemShut {NoStop}%
\bibitem [{\citenamefont {Gurvitz}\ and\ \citenamefont
  {Prager}(1996)}]{Gurvitz96:15932}%
  \BibitemOpen
  \bibfield  {author} {\bibinfo {author} {\bibfnamefont {S.~A.}\ \bibnamefont
  {Gurvitz}}\ and\ \bibinfo {author} {\bibfnamefont {Y.~S.}\ \bibnamefont
  {Prager}},\ }\href@noop {} {\bibfield  {journal} {\bibinfo  {journal} {Phys.
  Rev. B},\ }\textbf {\bibinfo {volume} {53}},\ \bibinfo {pages} {15932}
  (\bibinfo {year} {1996})}\BibitemShut {NoStop}%
\bibitem [{\citenamefont {van Kampen}(2001)}]{Kampen01:00}%
  \BibitemOpen
  \bibfield  {author} {\bibinfo {author} {\bibfnamefont {N.~G.}\ \bibnamefont
  {van Kampen}},\ }\href@noop {} {\emph {\bibinfo {title} {Stochastic Processes
  in Physics and Chemistry}}},\ \bibinfo {edition} {2nd}\ ed.\ (\bibinfo
  {publisher} {North-Holland, Amsterdam},\ \bibinfo {year} {2001})\BibitemShut
  {NoStop}%
\bibitem [{\citenamefont {Harbola}\ \emph {et~al.}(2006)\citenamefont
  {Harbola}, \citenamefont {Esposito},\ and\ \citenamefont
  {Mukamel}}]{Harbola06:235309}%
  \BibitemOpen
  \bibfield  {author} {\bibinfo {author} {\bibfnamefont {U.}~\bibnamefont
  {Harbola}}, \bibinfo {author} {\bibfnamefont {M.}~\bibnamefont {Esposito}}, \
  and\ \bibinfo {author} {\bibfnamefont {S.}~\bibnamefont {Mukamel}},\ }\href
  {http://link.aps.org/abstract/PRB/v74/e235309} {\bibfield  {journal}
  {\bibinfo  {journal} {Phys. Rev. B},\ }\textbf {\bibinfo {volume} {74}},\
  \bibinfo {pages} {235309} (\bibinfo {year} {2006})}\BibitemShut {NoStop}%
\bibitem [{\citenamefont {Braggio}\ \emph {et~al.}(2006)\citenamefont
  {Braggio}, \citenamefont {K\"{o}nig},\ and\ \citenamefont
  {Fazio}}]{Braggio06:026805}%
  \BibitemOpen
  \bibfield  {author} {\bibinfo {author} {\bibfnamefont {A.}~\bibnamefont
  {Braggio}}, \bibinfo {author} {\bibfnamefont {J.}~\bibnamefont {K\"{o}nig}},
  \ and\ \bibinfo {author} {\bibfnamefont {R.}~\bibnamefont {Fazio}},\
  }\href@noop {} {\bibfield  {journal} {\bibinfo  {journal} {Phys. Rev.
  Lett.},\ }\textbf {\bibinfo {volume} {96}},\ \bibinfo {pages} {026805}
  (\bibinfo {year} {2006})}\BibitemShut {NoStop}%
\bibitem [{\citenamefont {Emary}\ \emph {et~al.}(2007)\citenamefont {Emary},
  \citenamefont {Marcos}, \citenamefont {Aguado},\ and\ \citenamefont
  {Brandes}}]{Emary07:161404}%
  \BibitemOpen
  \bibfield  {author} {\bibinfo {author} {\bibfnamefont {C.}~\bibnamefont
  {Emary}}, \bibinfo {author} {\bibfnamefont {D.}~\bibnamefont {Marcos}},
  \bibinfo {author} {\bibfnamefont {R.}~\bibnamefont {Aguado}}, \ and\ \bibinfo
  {author} {\bibfnamefont {T.}~\bibnamefont {Brandes}},\ }\Doi
  {10.1103/PhysRevB.76.161404} {\bibfield  {journal} {\bibinfo  {journal}
  {Phys. Rev. B},\ }\textbf {\bibinfo {volume} {76}},\ \bibinfo {pages}
  {161404(R)} (\bibinfo {year} {2007})}\BibitemShut {NoStop}%
\bibitem [{\citenamefont {Bednorz}\ and\ \citenamefont
  {Belzig}(2008)}]{Bednorz08:206803}%
  \BibitemOpen
  \bibfield  {author} {\bibinfo {author} {\bibfnamefont {A.}~\bibnamefont
  {Bednorz}}\ and\ \bibinfo {author} {\bibfnamefont {W.}~\bibnamefont
  {Belzig}},\ }\Doi {10.1103/PhysRevLett.101.206803} {\bibfield  {journal}
  {\bibinfo  {journal} {Phys. Rev. Lett.},\ }\textbf {\bibinfo {volume}
  {101}},\ \bibinfo {pages} {206803} (\bibinfo {year} {2008})}\BibitemShut
  {NoStop}%
\bibitem [{\citenamefont {Moldoveanu}\ \emph {et~al.}(2009)\citenamefont
  {Moldoveanu}, \citenamefont {Manolescu},\ and\ \citenamefont
  {Gudmundsson}}]{Moldoveanu09:073019}%
  \BibitemOpen
  \bibfield  {author} {\bibinfo {author} {\bibfnamefont {V.}~\bibnamefont
  {Moldoveanu}}, \bibinfo {author} {\bibfnamefont {A.}~\bibnamefont
  {Manolescu}}, \ and\ \bibinfo {author} {\bibfnamefont {V.}~\bibnamefont
  {Gudmundsson}},\ }\href {http://stacks.iop.org/1367-2630/11/073019}
  {\bibfield  {journal} {\bibinfo  {journal} {New Journal of Physics},\
  }\textbf {\bibinfo {volume} {11}},\ \bibinfo {pages} {073019} (\bibinfo
  {year} {2009})}\BibitemShut {NoStop}%
\bibitem [{\citenamefont {Gudmundsson}\ \emph {et~al.}(2009)\citenamefont
  {Gudmundsson}, \citenamefont {Gainar}, \citenamefont {Tang}, \citenamefont
  {Moldoveanu},\ and\ \citenamefont {Manolescu}}]{Gudmundsson09:113007}%
  \BibitemOpen
  \bibfield  {author} {\bibinfo {author} {\bibfnamefont {V.}~\bibnamefont
  {Gudmundsson}}, \bibinfo {author} {\bibfnamefont {C.}~\bibnamefont {Gainar}},
  \bibinfo {author} {\bibfnamefont {C.-S.}\ \bibnamefont {Tang}}, \bibinfo
  {author} {\bibfnamefont {V.}~\bibnamefont {Moldoveanu}}, \ and\ \bibinfo
  {author} {\bibfnamefont {A.}~\bibnamefont {Manolescu}},\ }\href
  {http://stacks.iop.org/1367-2630/11/113007} {\bibfield  {journal} {\bibinfo
  {journal} {New Journal of Physics},\ }\textbf {\bibinfo {volume} {11}},\
  \bibinfo {pages} {113007} (\bibinfo {year} {2009})}\BibitemShut {NoStop}%
\bibitem [{\citenamefont {Vaz}\ and\ \citenamefont
  {Kyriakidis}(2008)}]{Vaz08:012012}%
  \BibitemOpen
  \bibfield  {author} {\bibinfo {author} {\bibfnamefont {E.}~\bibnamefont
  {Vaz}}\ and\ \bibinfo {author} {\bibfnamefont {J.}~\bibnamefont
  {Kyriakidis}},\ }\href@noop {} {\bibfield  {journal} {\bibinfo  {journal}
  {Journal of Physics Conference Series},\ }\textbf {\bibinfo {volume} {107}},\
  \bibinfo {pages} {012012} (\bibinfo {year} {2008})}\BibitemShut {NoStop}%
\bibitem [{\citenamefont {Gudmundsson}\ \emph {et~al.}(2010)\citenamefont
  {Gudmundsson}, \citenamefont {Tang}, \citenamefont {Jonasson}, \citenamefont
  {Moldoveanu},\ and\ \citenamefont {Manolescu}}]{Gudmundsson10:205319}%
  \BibitemOpen
  \bibfield  {author} {\bibinfo {author} {\bibfnamefont {V.}~\bibnamefont
  {Gudmundsson}}, \bibinfo {author} {\bibfnamefont {C.-S.}\ \bibnamefont
  {Tang}}, \bibinfo {author} {\bibfnamefont {O.}~\bibnamefont {Jonasson}},
  \bibinfo {author} {\bibfnamefont {V.}~\bibnamefont {Moldoveanu}}, \ and\
  \bibinfo {author} {\bibfnamefont {A.}~\bibnamefont {Manolescu}},\ }\href
  {http://link.aps.org/doi/10.1103/PhysRevB.81.205319} {\bibfield  {journal}
  {\bibinfo  {journal} {Phys. Rev. B},\ }\textbf {\bibinfo {volume} {81}},\
  \bibinfo {pages} {205319} (\bibinfo {year} {2010})}\BibitemShut {NoStop}%
\bibitem [{\citenamefont {Helm}(2000)}]{Helm00:01}%
  \BibitemOpen
  \bibfield  {author} {\bibinfo {author} {\bibfnamefont {M.}~\bibnamefont
  {Helm}},\ }\href@noop {} {\emph {\bibinfo {title} {Intersubband Transitions
  in Quantum Wells: Physics and Device Applications I}}},\ edited by\ \bibinfo
  {editor} {\bibfnamefont {H.~C.}\ \bibnamefont {Liu}}\ and\ \bibinfo {editor}
  {\bibfnamefont {F.}~\bibnamefont {Capasso}}\ (\bibinfo  {publisher} {Academic
  Press},\ \bibinfo {year} {2000})\BibitemShut {NoStop}%
\bibitem [{\citenamefont {Gabbay}\ \emph {et~al.}(2011)\citenamefont {Gabbay},
  \citenamefont {Reno}, \citenamefont {Wendt}, \citenamefont {Gin},
  \citenamefont {Wanke}, \citenamefont {Sinclair}, \citenamefont {Shaner},\
  and\ \citenamefont {Brener}}]{Gabbay11:203103}%
  \BibitemOpen
  \bibfield  {author} {\bibinfo {author} {\bibfnamefont {A.}~\bibnamefont
  {Gabbay}}, \bibinfo {author} {\bibfnamefont {J.}~\bibnamefont {Reno}},
  \bibinfo {author} {\bibfnamefont {J.~R.}\ \bibnamefont {Wendt}}, \bibinfo
  {author} {\bibfnamefont {A.}~\bibnamefont {Gin}}, \bibinfo {author}
  {\bibfnamefont {M.~C.}\ \bibnamefont {Wanke}}, \bibinfo {author}
  {\bibfnamefont {M.~B.}\ \bibnamefont {Sinclair}}, \bibinfo {author}
  {\bibfnamefont {E.}~\bibnamefont {Shaner}}, \ and\ \bibinfo {author}
  {\bibfnamefont {I.}~\bibnamefont {Brener}},\ }\Doi {doi:10.1063/1.3592266}
  {\bibfield  {journal} {\bibinfo  {journal} {Appl. Phys. Lett.},\ }\textbf
  {\bibinfo {volume} {98}},\ \bibinfo {pages} {203103} (\bibinfo {year}
  {2011})}\BibitemShut {NoStop}%
\bibitem [{\citenamefont {Ciuti}\ \emph {et~al.}(2005)\citenamefont {Ciuti},
  \citenamefont {Bastard},\ and\ \citenamefont {Carusotto}}]{Ciuti05:115303}%
  \BibitemOpen
  \bibfield  {author} {\bibinfo {author} {\bibfnamefont {C.}~\bibnamefont
  {Ciuti}}, \bibinfo {author} {\bibfnamefont {G.}~\bibnamefont {Bastard}}, \
  and\ \bibinfo {author} {\bibfnamefont {I.}~\bibnamefont {Carusotto}},\ }\Doi
  {10.1103/PhysRevB.72.115303} {\bibfield  {journal} {\bibinfo  {journal}
  {Phys. Rev. B},\ }\textbf {\bibinfo {volume} {72}},\ \bibinfo {pages}
  {115303} (\bibinfo {year} {2005})}\BibitemShut {NoStop}%
\bibitem [{\citenamefont {Devoret}\ \emph {et~al.}(2007)\citenamefont
  {Devoret}, \citenamefont {Girvin},\ and\ \citenamefont
  {Schoelkopf}}]{Devoret07:767}%
  \BibitemOpen
  \bibfield  {author} {\bibinfo {author} {\bibfnamefont {M.}~\bibnamefont
  {Devoret}}, \bibinfo {author} {\bibfnamefont {S.}~\bibnamefont {Girvin}}, \
  and\ \bibinfo {author} {\bibfnamefont {R.}~\bibnamefont {Schoelkopf}},\
  }\href
  {http://qulab.eng.yale.edu/documents/papers/Devoret,Girvin,Schoelkopf-circuitQEDcouplingStrength-AnnPhys16,767(Oct2007).pdf}
  {\bibfield  {journal} {\bibinfo  {journal} {Ann. Phys.},\ }\textbf {\bibinfo
  {volume} {16}},\ \bibinfo {pages} {767} (\bibinfo {year} {2007})}\BibitemShut
  {NoStop}%
\bibitem [{\citenamefont {Abdumalikov}\ \emph {et~al.}(2008)\citenamefont
  {Abdumalikov}, \citenamefont {Astafiev}, \citenamefont {Nakamura},
  \citenamefont {Pashkin},\ and\ \citenamefont {Tsai}}]{Abdumalikov08:180502}%
  \BibitemOpen
  \bibfield  {author} {\bibinfo {author} {\bibfnamefont {A.~A.}\ \bibnamefont
  {Abdumalikov}}, \bibinfo {author} {\bibfnamefont {O.}~\bibnamefont
  {Astafiev}}, \bibinfo {author} {\bibfnamefont {Y.}~\bibnamefont {Nakamura}},
  \bibinfo {author} {\bibfnamefont {Y.~A.}\ \bibnamefont {Pashkin}}, \ and\
  \bibinfo {author} {\bibfnamefont {J.}~\bibnamefont {Tsai}},\ }\Doi
  {10.1103/PhysRevB.78.180502} {\bibfield  {journal} {\bibinfo  {journal}
  {Phys. Rev. B},\ }\textbf {\bibinfo {volume} {78}},\ \bibinfo {pages}
  {180502(R)} (\bibinfo {year} {2008})}\BibitemShut {NoStop}%
\bibitem [{\citenamefont {Zela}\ \emph {et~al.}(1997)\citenamefont {Zela},
  \citenamefont {Solano},\ and\ \citenamefont {Gago}}]{Zela97:106}%
  \BibitemOpen
  \bibfield  {author} {\bibinfo {author} {\bibfnamefont {F.~D.}\ \bibnamefont
  {Zela}}, \bibinfo {author} {\bibfnamefont {E.}~\bibnamefont {Solano}}, \ and\
  \bibinfo {author} {\bibfnamefont {A.}~\bibnamefont {Gago}},\ }\Doi
  {10.1016/S0030-4018(97)00263-0} {\bibfield  {journal} {\bibinfo  {journal}
  {Optics Communications},\ }\textbf {\bibinfo {volume} {142}},\ \bibinfo
  {pages} {106} (\bibinfo {year} {1997})}\BibitemShut {NoStop}%
\bibitem [{\citenamefont {Sornborger}\ and\ \citenamefont
  {Geller}(2004)}]{Sornborger04:052315}%
  \BibitemOpen
  \bibfield  {author} {\bibinfo {author} {\bibfnamefont {A.~T.}\ \bibnamefont
  {Sornborger}}\ and\ \bibinfo {author} {\bibfnamefont {A.~N. C. M.~R.}\
  \bibnamefont {Geller}},\ }\href@noop {} {\bibfield  {journal} {\bibinfo
  {journal} {Phys. Rev. A},\ }\textbf {\bibinfo {volume} {70}},\ \bibinfo
  {pages} {052315} (\bibinfo {year} {2004})}\BibitemShut {NoStop}%
\bibitem [{\citenamefont {Irish}(2007)}]{Irish07:173601}%
  \BibitemOpen
  \bibfield  {author} {\bibinfo {author} {\bibfnamefont {E.~K.}\ \bibnamefont
  {Irish}},\ }\Doi {10.1103/PhysRevLett.99.173601} {\bibfield  {journal}
  {\bibinfo  {journal} {Phys. Rev. Lett.},\ }\textbf {\bibinfo {volume} {99}},\
  \bibinfo {pages} {173601} (\bibinfo {year} {2007})}\BibitemShut {NoStop}%
\bibitem [{\citenamefont {Delbecq}\ \emph {et~al.}(2011)\citenamefont
  {Delbecq}, \citenamefont {Schmitt}, \citenamefont {Parmentier}, \citenamefont
  {Roch}, \citenamefont {Viennot}, \citenamefont {F{\`e}ve}, \citenamefont
  {Huard}, \citenamefont {Mora}, \citenamefont {Cottet},\ and\ \citenamefont
  {Kontos}}]{Delbecq11:01}%
  \BibitemOpen
  \bibfield  {author} {\bibinfo {author} {\bibfnamefont {M.}~\bibnamefont
  {Delbecq}}, \bibinfo {author} {\bibfnamefont {V.}~\bibnamefont {Schmitt}},
  \bibinfo {author} {\bibfnamefont {F.}~\bibnamefont {Parmentier}}, \bibinfo
  {author} {\bibfnamefont {N.}~\bibnamefont {Roch}}, \bibinfo {author}
  {\bibfnamefont {J.}~\bibnamefont {Viennot}}, \bibinfo {author} {\bibfnamefont
  {G.}~\bibnamefont {F{\`e}ve}}, \bibinfo {author} {\bibfnamefont
  {B.}~\bibnamefont {Huard}}, \bibinfo {author} {\bibfnamefont
  {C.}~\bibnamefont {Mora}}, \bibinfo {author} {\bibfnamefont {A.}~\bibnamefont
  {Cottet}}, \ and\ \bibinfo {author} {\bibfnamefont {T.}~\bibnamefont
  {Kontos}},\ }\href {http://arxiv.org/abs/1108.4371} {\bibfield  {journal}
  {\bibinfo  {journal} {arXiv:1108.4371}} (\bibinfo {year} {2011})}\BibitemShut
  {NoStop}%
\bibitem [{\citenamefont {Frey}\ \emph {et~al.}(2011)\citenamefont {Frey},
  \citenamefont {Leek}, \citenamefont {Beck}, \citenamefont {Blais},
  \citenamefont {Ihn}, \citenamefont {Ensslin},\ and\ \citenamefont
  {Wallraff}}]{Frey11:01}%
  \BibitemOpen
  \bibfield  {author} {\bibinfo {author} {\bibfnamefont {T.}~\bibnamefont
  {Frey}}, \bibinfo {author} {\bibfnamefont {P.~J.}\ \bibnamefont {Leek}},
  \bibinfo {author} {\bibfnamefont {M.}~\bibnamefont {Beck}}, \bibinfo {author}
  {\bibfnamefont {A.}~\bibnamefont {Blais}}, \bibinfo {author} {\bibfnamefont
  {T.}~\bibnamefont {Ihn}}, \bibinfo {author} {\bibfnamefont {K.}~\bibnamefont
  {Ensslin}}, \ and\ \bibinfo {author} {\bibfnamefont {A.}~\bibnamefont
  {Wallraff}},\ }\href {http://arxiv.org/abs/1108.5378} {\bibfield  {journal}
  {\bibinfo  {journal} {arXiv:1108.5378}} (\bibinfo {year} {2011})}\BibitemShut
  {NoStop}%
\bibitem [{\citenamefont {Nakajima}(1958)}]{Nakajima58:948}%
  \BibitemOpen
  \bibfield  {author} {\bibinfo {author} {\bibfnamefont {S.}~\bibnamefont
  {Nakajima}},\ }\href@noop {} {\bibfield  {journal} {\bibinfo  {journal}
  {Prog. Theor. Phys.},\ }\textbf {\bibinfo {volume} {20}},\ \bibinfo {pages}
  {948} (\bibinfo {year} {1958})}\BibitemShut {NoStop}%
\bibitem [{\citenamefont {Zwanzig}(1960)}]{Zwanzig60:1338}%
  \BibitemOpen
  \bibfield  {author} {\bibinfo {author} {\bibfnamefont {R.}~\bibnamefont
  {Zwanzig}},\ }\href@noop {} {\bibfield  {journal} {\bibinfo  {journal} {J.
  Chem. Phys.},\ }\textbf {\bibinfo {volume} {33}},\ \bibinfo {pages} {1338}
  (\bibinfo {year} {1960})}\BibitemShut {NoStop}%
\bibitem [{\citenamefont {Moldoveanu}\ \emph {et~al.}(2010)\citenamefont
  {Moldoveanu}, \citenamefont {Manolescu}, \citenamefont {Tang},\ and\
  \citenamefont {Gudmundsson}}]{Moldoveanu10:155442}%
  \BibitemOpen
  \bibfield  {author} {\bibinfo {author} {\bibfnamefont {V.}~\bibnamefont
  {Moldoveanu}}, \bibinfo {author} {\bibfnamefont {A.}~\bibnamefont
  {Manolescu}}, \bibinfo {author} {\bibfnamefont {C.-S.}\ \bibnamefont {Tang}},
  \ and\ \bibinfo {author} {\bibfnamefont {V.}~\bibnamefont {Gudmundsson}},\
  }\href {http://link.aps.org/doi/10.1103/PhysRevB.81.155442} {\bibfield
  {journal} {\bibinfo  {journal} {Phys. Rev. B},\ }\textbf {\bibinfo {volume}
  {81}},\ \bibinfo {pages} {155442} (\bibinfo {year} {2010})}\BibitemShut
  {NoStop}%
\bibitem [{\citenamefont {Jonasson}(2011)}]{Jonasson2011:01}%
  \BibitemOpen
  \bibfield  {author} {\bibinfo {author} {\bibfnamefont {O.}~\bibnamefont
  {Jonasson}},\ }\href {http://arxiv.org/abs/1109.4594} {\bibfield  {journal}
  {\bibinfo  {journal} {arXiv:1109.4594}} (\bibinfo {year} {2011})}\BibitemShut
  {NoStop}%
\bibitem [{\citenamefont {Magn{\'u}sd{\'o}ttir}\ and\ \citenamefont
  {Gudmundsson}(1999)}]{Ingibjorg99:16591}%
  \BibitemOpen
  \bibfield  {author} {\bibinfo {author} {\bibfnamefont {I.}~\bibnamefont
  {Magn{\'u}sd{\'o}ttir}}\ and\ \bibinfo {author} {\bibfnamefont
  {V.}~\bibnamefont {Gudmundsson}},\ }\href@noop {} {\bibfield  {journal}
  {\bibinfo  {journal} {Phys. Rev. B},\ }\textbf {\bibinfo {volume} {60}},\
  \bibinfo {pages} {16591} (\bibinfo {year} {1999})}\BibitemShut {NoStop}%
\bibitem [{\citenamefont {Magn{\'u}sd{\'o}ttir}\ and\ \citenamefont
  {Gudmundsson}(2000)}]{Magnusdottir00:10229}%
  \BibitemOpen
  \bibfield  {author} {\bibinfo {author} {\bibfnamefont {I.}~\bibnamefont
  {Magn{\'u}sd{\'o}ttir}}\ and\ \bibinfo {author} {\bibfnamefont
  {V.}~\bibnamefont {Gudmundsson}},\ }\href@noop {} {\bibfield  {journal}
  {\bibinfo  {journal} {Phys. Rev. B},\ }\textbf {\bibinfo {volume} {61}},\
  \bibinfo {pages} {10229} (\bibinfo {year} {2000})}\BibitemShut {NoStop}%
\bibitem [{\citenamefont {Feranchuk}\ \emph {et~al.}(1996)\citenamefont
  {Feranchuk}, \citenamefont {Komarov},\ and\ \citenamefont
  {Ulyanenkov}}]{Feranchuk96:4035}%
  \BibitemOpen
  \bibfield  {author} {\bibinfo {author} {\bibfnamefont {I.~D.}\ \bibnamefont
  {Feranchuk}}, \bibinfo {author} {\bibfnamefont {L.~I.}\ \bibnamefont
  {Komarov}}, \ and\ \bibinfo {author} {\bibfnamefont {A.~P.}\ \bibnamefont
  {Ulyanenkov}},\ }\href {http://iopscience.iop.org/0305-4470/29/14/026}
  {\bibfield  {journal} {\bibinfo  {journal} {J. Phys. A: Math. Gen.},\
  }\textbf {\bibinfo {volume} {29}},\ \bibinfo {pages} {4035} (\bibinfo {year}
  {1996})}\BibitemShut {NoStop}%
\bibitem [{\citenamefont {Li}\ \emph {et~al.}(2009)\citenamefont {Li},
  \citenamefont {Wang},\ and\ \citenamefont {Liu}}]{Li09:044212}%
  \BibitemOpen
  \bibfield  {author} {\bibinfo {author} {\bibfnamefont {X.-H.}\ \bibnamefont
  {Li}}, \bibinfo {author} {\bibfnamefont {K.-L.}\ \bibnamefont {Wang}}, \ and\
  \bibinfo {author} {\bibfnamefont {T.}~\bibnamefont {Liu}},\ }\href
  {http://iopscience.iop.org/0256-307X/26/4/044212} {\bibfield  {journal}
  {\bibinfo  {journal} {Chin. Phys. Lett.},\ }\textbf {\bibinfo {volume}
  {26}},\ \bibinfo {pages} {044212} (\bibinfo {year} {2009})}\BibitemShut
  {NoStop}%
\bibitem [{\citenamefont {Jin}\ \emph {et~al.}(2011)\citenamefont {Jin},
  \citenamefont {Li}, \citenamefont {Luo},\ and\ \citenamefont
  {Yan}}]{Jin11:053704}%
  \BibitemOpen
  \bibfield  {author} {\bibinfo {author} {\bibfnamefont {J.}~\bibnamefont
  {Jin}}, \bibinfo {author} {\bibfnamefont {X.-Q.}\ \bibnamefont {Li}},
  \bibinfo {author} {\bibfnamefont {M.}~\bibnamefont {Luo}}, \ and\ \bibinfo
  {author} {\bibfnamefont {Y.}~\bibnamefont {Yan}},\ }\Doi
  {doi:10.1063/1.3555586} {\bibfield  {journal} {\bibinfo  {journal} {J. Appl.
  Phys.},\ }\textbf {\bibinfo {volume} {109}},\ \bibinfo {pages} {053704}
  (\bibinfo {year} {2011})}\BibitemShut {NoStop}%
\bibitem [{\citenamefont {Whitney}(2008)}]{Whitney08:175304}%
  \BibitemOpen
  \bibfield  {author} {\bibinfo {author} {\bibfnamefont {R.~S.}\ \bibnamefont
  {Whitney}},\ }\Doi {10.1088/1751-8113/41/17/175304} {\bibfield  {journal}
  {\bibinfo  {journal} {J. Phys. A: Math. Theor.},\ }\textbf {\bibinfo {volume}
  {41}},\ \bibinfo {pages} {175304} (\bibinfo {year} {2008})}\BibitemShut
  {NoStop}%
\end{thebibliography}
%

%
%
%
\end{document}